\documentclass[a4paper,12pt,useAMS]{mn2e}

\usepackage{graphicx}

\title[Mass Models \& Einstein Radii of the 12 $\lowercase{z}>0.5$ MACS clusters]{Strong-Lensing Analysis of
a Complete Sample of 12 MACS Clusters at $\lowercase{z}>0.5$: Mass Models and Einstein Radii}

\author[Zitrin et al.]{Adi Zitrin$^{1}$\thanks{E-mail:
adiz@wise.tau.ac.il}, Tom Broadhurst$^{1,2,3}$, Rennan Barkana$^{1}$, Yoel Rephaeli$^{1}$, Narciso Ben\'{\i}tez$^{4}$\\\\\\
$^{1}$The School of Physics and Astronomy, the Raymond and Beverly Sackler Faculty of Exact Sciences, Tel Aviv University,\\ Tel Aviv 69978, Israel\\
$^{2}$Department of Theoretical Physics, University of Basque Country UPV/EHU, Leioa, Spain\\
$^{3}$IKERBASQUE, Basque Foundation for Science, 48011, Bilbao, Spain\\
$^{4}$Instituto de Astrof\'isica de Andaluc\'ia (CSIC), C/Camino Bajo de Hu\'etor, 24, Granada, 18008, Spain}

\begin{document}

%\date{Accepted 1988 December 15. Received 1988 December 14; in original form 1988 October 11}

\pagerange{\pageref{firstpage}--\pageref{lastpage}} \pubyear{2010}

\maketitle

\label{firstpage}

\begin{abstract}

We present the results of a strong-lensing analysis
of a complete sample of 12 very luminous X-ray clusters at $z>0.5$ using HST/ACS
images. Our modelling technique has uncovered some of the largest known critical curves outlined by many accurately-predicted sets of multiple images. The
distribution of Einstein radii has a median value of $\simeq28\arcsec$ (for a source
redshift of $z_{s}\sim2$), twice as large as other lower-$z$ samples, and extends to $55\arcsec$ for MACS J0717.5+3745, with an impressive
enclosed Einstein mass of $7.4\times10^{14} M_{\odot}$. We find that 9 clusters cover a very large area ($>2.5 \sq \arcmin$) of high magnification ($\mu > \times10$) for a source redshift of $z_{s}\sim8$, providing primary targets for accessing the first stars and galaxies. We compare our results with theoretical predictions of the standard $\Lambda$CDM model which we show systematically fall short of our measured Einstein radii by a factor of $\simeq1.4$, after accounting for the effect of lensing projection. Nevertheless, a revised analysis once arc redshifts become available, and similar analyses of larger samples, are needed in order to establish more precisely the level of discrepancy with $\Lambda$CDM predictions.

\end{abstract}


\section{Introduction}

Massive galaxy clusters provide several means for independently
examining any viable cosmological model. Cluster samples used for this
purpose are usually complete in terms of X-ray and redshift
measurements such as the RDCS (Rosati et al. 1998, Borgani et al. 1999) and the MACS
survey (Ebeling, Edge, \& Henry 2001, Ebeling et al. 2007) that we examine here with lensing. Ideally, controlled samples of clusters with reliable lensing masses would allow the most direct comparison with theory but these are presently still in their infancy with only a handful of clusters discovered in wide-field weak lensing searches (Taylor et al. 2004, Wittman et al. 2006, Massey et al. 2007, Miyazaki et al. 2007), and may be often subject to projection biases (e.g., Hennawi \& Spergel 2005). Weak lensing surveys should have the advantage of being approximately volume-limited over the redshift range 0.1<$z$<1.0, where the lensing distance ratio has a broad maximum.

Ongoing measurements of the Sunyaev-Zel'dovich (SZ) effect (which
has already been detected in many tens of clusters, including some at
relatively high redshifts $z\sim 0.9$, e.g., LaRoque et al. 2006) in a
large number of clusters over a wide redshift range should improve upon X-ray
selection, due to the redshift-independence of the effect (though clearly high
redshift clusters are relatively less well resolved). Also, since the insensitivity of the effect
to internal gas density variations will not be biased as with X-ray
flux selection towards systems with luminous shocked gas,
measurements should better correlate with cluster masses.

Strong gravitational lensing (SL) is nearly always seen in
sufficiently detailed observations of massive clusters. Einstein radii
derived from such observations provide a reliable projected mass
within the observed Einstein radius, which depends only on fundamental
constants, G and c, and knowledge of the lens and source distances.
Constraining the inner mass profile from SL is much harder, requiring
many sets of multiple images covering a wide range of source redshift,
to overcome inherent lensing degeneracy between the gradient of the
mass profile and the scaling of the bend angle with source distance,
so that SL based mass profiles have been usefully constrained for only
several clusters producing reliable mass maps (e.g., Abell 370, Kneib et al. 1993, Richard et al. 2009a; Abell 901, Deb et al. 2010; Abell 1689, Broadhurst et al. 2005, Coe et al. 2010; Abell 1703, Limousin et al. 2008; Cl0024+1654, Liesenborgs et al. 2008, Zitrin et al. 2009b; MS 2137.3-2353, Gavazzi et al. 2003, Merten et al. 2009; RXJ1347, Brada\v{c} et al. 2008a, Halkola et al. 2008; SDSS J1004+4112, Sharon et al. 2005, Liesenborgs et al. 2009; "The bullet cluster", Brada\v{c} et al. 2006).

In quite a few of these clusters there is a significant inconsistency between the Einstein radius directly measured from SL analyses and the Einstein radius predicted from model profiles fitted to weak-lensing measurements of the same clusters, often subject to background selection and dilution problems over a wide radial range (e.g., Medezinski et al. 2010), though more so towards the SL regime where the degree of contamination by cluster members is often poorly corrected. The importance of a SL analysis of a significant sample is clear, especially in the statistical aspect, where the modelling method is similar for all examined clusters and thus supplies a coherent view for objective comparison (e.g., Richard et al. 2009b).

The development of SL modelling-methods has increased in response to
the improvement in data quality since the discoveries of the first giant arcs (e.g., Lynds \& Petrosian 1986, Soucail et al. 1987, Kneib et al. 1996). Most methods can be classified as ``parametric'' if based on physical parameterisation, and as ``non-parametric'' if they are ``grid-based'' (see also \S4.4 in Coe et al. 2008, and references therein). Still, many of these methods include too many parameters to be
well-constrained by the number of initially known multiply-lensed systems. Here we use ACS imaging to identify new multiply-lensed systems in order to constrain the mass distributions and define the critical curves of the sample, motivated by the successful minimalistic approach of Broadhurst et al. (2005) to lens modelling. In following work (Zitrin et al. 2009b) we presented an improved modelling method which involved only 6 free parameters enabling easier constraint, since the number of constraints has to be equal or larger to the number of parameters in order to get a reliable fit. Two of these parameters are primarily set to reasonable values so only 4 of these parameters have to be constrained initially, which sets a very reliable starting-point using obvious systems. The mass distribution is therefore primarily well constrained, uncovering many multiple-images which can be then iteratively incorporated into the model, by using their redshift estimation and location in the image-plane.

The Massive Cluster Survey (MACS) has been aimed to create a complete sample of the very X-ray luminous clusters in the Universe, successfully
increasing the number
of known such clusters
to hundreds, particularly at $z>0.3$ (Ebeling, Edge, \& Henry 2001). From this a
complete sample of 12 high-$z$ MACS clusters ($z>0.5$) was defined by Ebeling
et al. (2007) which have proved very interesting in several follow-up studies
including deep X-ray, SZ and HST imaging. Detection of a large-scale filament has been reported in the case of MACS J0717.5+3745 by
Ebeling et al. (2004), for which many multiply-lensed images have been recently identified by Zitrin et al. (2009a), revealing this object to
be the largest known lens, with an Einstein radius equivalent to 55$\arcsec$ (for a source at $z\sim2.5$). In the case of MACS J1149.5+2223 (Zitrin \& Broadhurst 2009), a background spiral galaxy at $z=1.49$ (Smith et al. 2009) has been shown to be multiply-lensed into several very large images, requiring a very shallow, unrelaxed central mass distribution. Another large multiply-lensed sub-mm source at a redshift of $z\simeq2.9$ has been identified in MACS J0454.1-0300 (also referred to as MS 0451.6-0305; Takata et al. 2003, Borys et al. 2004, Berciano Alba et al. 2007, 2009), MACS J0025.4-1222 was found to be a ``bullet cluster''-like (Brada\v{c} et al. 2008b), and other MACS clusters have been recently used for an extensive arc statistics study (Horesh et al. 2010).

The X-ray data available for this sample (see Ebeling et al. 2007) along with the high-resolution HST/ACS imaging and
SZ imaging (e.g., Laroque et al. 2003)  make these MACS targets particularly useful for understanding the nature of the most massive clusters. Here we complete a full SL analysis of this 12 cluster sample with the available deep HST/ACS imaging, principally to derive their Einstein radii and
also to help constrain the central mass distributions. The effective Einstein radii derived here are the corresponding radii of circles of equivalent areas to those enclosed within the critical curves, or similarly, the radii within which the averaged $\kappa$ is equal to 1. In addition, this work can also supply detailed magnification maps to help motivate deeper searches for high-$z$ galaxies behind these clusters.

Another major motivation for pursuing accurate lensing-maps
is the increased precision of model predictions for cluster-
size massive halos in the standard $\Lambda$CDM model and close variance (see Umetsu \& Broadhurst 2008, e.g., Bullock et
al. 2001, Spergel et al. 2003, 2007, Tegmark et al. 2004, Hennawi et al. 2007, Neto et al. 2007, Duffy et
al. 2008, Keselman, Nusser \& Peebles 2009, Meneghetti et al. 2010, Fedeli et al. 2010). In the standard hierarchical model, large massive bodies are due to collapse recently and thus are not expected to be found in large numbers at high redshifts. On the other hand, the larger volume available with increasing distance means that in
practice we cannot expect to reside next to the most massive
cluster (see also Zitrin et al. 2009a). The increasing number of accurately analysed clusters have revealed larger lenses than predicted by the standard $\Lambda CDM$ model (Broadhurst \& Barkana 2008, Puchwein \& Hilbert 2009). This discrepancy is empirically supported by the surprisingly concentrated mass profiles measured for such clusters, when combining the inner strong lensing with the outer weak lensing signal (Gavazzi et al. 2003, Broadhurst et al. 2005 \& 2008, Limousin et al. 2008, Donnarumma et al. 2009, Oguri et al. 2009, Umetsu et al. 2009, Zitrin et al. 2009b,2010), or independently, when using the internal dynamics of cluster members (Lemze et al. 2009) and deep X-ray data (Lemze et al. 2008). Geometrically, lensing is considered to be optimised at intermediate redshifts,
where for a given mass the critical density for lensing is minimal, but
this is partly offset by the late hierarchical growth of high-mass systems. This trade-off results in estimates
of the amplitude of strong lensing to favour the redshift range
$z=0.2-0.4$. Our recent analysis of two X-ray selected MACS clusters reveals very large lenses (e.g.,  MACS J1149.5+2223, Zitrin \& Broadhurst 2009; MACS 0717.5+3745, Zitrin et al. 2009a) at high redshift, increasing
the tension with $\Lambda$CDM predictions, since large Einstein radius clusters
at high-$z$ require earlier formation
than implied by the standard $\Lambda$CDM model (Sadeh \& Rephaeli 2008). The
largest lensing clusters have proven to be excellent targets for accessing the
faint early Universe due to their large magnification consistently providing
the highest redshift galaxies (Ebbels et al. 1996, Franx et al. 1997, Frye \&
Broadhurst 1998, Bouwens et al. 2004, Kneib et al. 2004, Bradley et al. 2008,
Zheng et al. 2009).

The statistics of large Einstein radii provide an important opportunity to test the standard $\Lambda$CDM paradigm, as it probes both the high-mass end of the cluster mass function and central mass distributions of massive clusters (Oguri \& Blandford 2009). Increasing numbers of theoretical and observational works have been done recently, deriving Einstein radius distributions of various samples. For example, Cypriano et al. (2003) have derived the mass distributions of 24 X-ray selected Abell clusters via weak-lensing. By fitting to SIS profiles they obtain a mean Einstein radius of $\sim17 \arcsec$. More recently, Hoekstra (2007) finds by weak-lensing analysis and SIS fitting a mean Einstein radius of $\sim14 \arcsec$ for a sample of 20 X-ray luminous clusters. Okabe et al. (2009) find the mass distribution and Einstein radii for a sample of $\sim30$ LoCuSS clusters by fitting the weak-lensing data to CIS profiles, obtaining a mean value of $\sim22 \arcsec$ (see Okabe et al. 2009 and references therein). Note that these values quoted here are calculated from the corresponding tables in these papers, and are not necessarily scaled or normalised to certain lens and source redshifts.

Other recent work more relevant to our study summarises SL analysis for a sample of 20, mostly relaxed \emph{undisturbed} clusters (Richard et al. 2009b). They find that the Einstein radii are distributed log-normally with a peak at $\theta_{e}=14.45 \arcsec$ and a corresponding $1.95 \times 10^{14} M_{\odot}$ enclosed mass (within R$<$250 kpc), and show that the predicted distribution of Einstein radii from $\Lambda$CDM cosmology falls short of the observed Einstein radii by a factor of 2. It has been claimed that much larger Einstein radii can be contemplated only with mass distributions which are highly prolate and aligned along the line of sight (Corless \& King 2007, Oguri \& Blandford 2009, see also Hennawi et al. 2007, Meneghetti et al. 2007, Sereno, Jetzer \& Lubini 2010). Analysing the 12 X-ray selected, high-$z$ MACS clusters is important since they are predicted to be massive and should form efficient lenses, whose properties can be then compared to such studies, playing an important role in probing the $\Lambda$CDM scenario.

The paper is organised as follows: in \S2 we describe the sample and observations; the lensing
analysis is described in \S3 detailing each cluster separately. Our
overall results are presented and discussed in \S4, followed by a Conclusion (\S5).
Throughout the paper we adopt the standard cosmology ($\Omega_{\rm m0}=0.3$,
$\Omega_{\Lambda0}=0.7$,$h=0.7$).

\section{The Sample and Observations}

The $z>0.5$ MACS clusters have been imaged with the Wide Field Channel
(WFC) of the ACS installed on HST, mainly in the framework of proposal
ID 9722 (PI: Ebeling). This mainly comprises two-filter observations
in the F555W and F814W bands, taken during 2003 and 2004 with typical
exposure times of $\sim 4500$s in each filter. We make use of some other
HST data available in the Hubble Legacy Archive (HLA), which we list
in Table \ref{sample}. In addition, available SExtractor (Bertin \& Arnouts
1996) photometry catalogues were also downloaded from the HLA, and were used
to construct colour-magnitude diagrams for identifying the red cluster sequence
galaxies belonging to each cluster, as a starting point for each
lensing model, as detailed in \S \ref{model}.
\begin{table*}
\caption{Properties of the MACS $ z>0.5$ sample. The following data (columns 1-7) are based on Ebeling et al. (2007): \emph{Column 1:} cluster name in the MACS survey; \emph{Columns 2 \& 3:} The RA and Declination of the X-ray centroids (as determined from Chandra ACIS-I data); \emph{Column 4:} redshifts; \emph{Column 5:} velocity dispersions, in km s$^{-1}$; \emph{Column 6:} Chandra X-ray luminosities in $10^{44}$ erg $s^{-1}$ quoted for the $0.1$-$2.4$ keV band. These luminosities are quoted within $r_{200}$ and exclude X-ray point sources; \emph{Column 7:} X-ray temperatures, measured from Chandra data within $r_{1000}$, but excluding a central region of 70 kpc radius around the listed X-ray centroid; \emph{Column 8:} Ebeling et al. (2007) morphology code - assessed visually based on the appearance of the X-ray contours and the goodness of the optical/X-ray alignment. The assigned codes (from apparently relaxed to extremely disturbed) are 1 (pronounced cool core, perfect alignment of X-ray peak and single cD galaxy), 2 (good optical/X-ray alignment, concentric contours), 3 (nonconcentric contours, obvious small-scale substructure), and 4 (poor optical/X-ray alignment, multiple peaks, no cD galaxy), errors are estimated as less than 1; \emph{Column 9:} ACS bands used here. Note, the clusters MACS J0018.5+1626 and MACS
J0454.1-0300 are better known from earlier work as Cl0016+1609 and
MS 0451.6-0305, respectively. For more information see Ebeling et
al. (2007).}
%\vspace{0.5cm}
\label{sample}
%\begin{footnotesize}
\begin{center}
\begin{tabular}{|c|c|c|c|c|c|c|c|c|}
  \hline\hline
  MACS & $\alpha$ (J2000.0) & $\delta$ (J2000.0) & $z$ & $\sigma$ & $L_{x, Chandra}$ & $KT (kev)$ & M.C.E. & ACS bands\\
  \hline
  J0018.5+1626 & 00 18 33.835 & +16 26 16.64 & 0.545 & $1420^{+140}_{-150}$ & $19.6\pm{0.3}$ & $9.4\pm{1.3}$ & 3 & F606W,F775W,F850LP\\
  J0025.4-1222 & 00 25 29.381 & -12 22 37.06 & 0.584 & $740^{+90}_{-110}$ & $8.8\pm{0.2}$  & $7.1\pm{0.7}$ & 3 & F450W,F555W,F814W\\
  J0257.1-2325 & 02 57 09.151 & -23 26 05.83 & 0.505 & $970^{+200}_{-250}$ & $ 13.7\pm{0.3}$ & $10.5\pm{1.0}$&  2 & F555W,F814W\\
  J0454.1-0300 & 04 54 11.125 & -03 00 53.77 & 0.538 & $1250^{+130}_{-180}$ & $16.8\pm{0.6}$ & $7.5\pm{1.0}$ & 2 & F555W,F775W,F814W\\
  J0647.7+7015 & 06 47 50.469 & +70 14 54.95 & 0.591 & $900^{+120}_{-180}$ & $15.9\pm{0.4}$ & $11.5\pm{1.0}$&  2 & F555W,F814W\\
  J0717.5+3745 & 07 17 30.927 & +37 45 29.74 & 0.546 & $1660^{+120 }_{-130}$ & $24.6\pm{0.3}$ & $11.6\pm{0.5}$&  4 & F555W,F606W,F814W\\
  J0744.8+3927 & 07 44 52.470 & +39 27 27.34 & 0.698 & $1110^{+130 }_{-150}$ & $22.9\pm{0.6}$ & $8.1\pm{0.6}$&  2 & F555W,F814W\\
  J0911.2+1746 & 09 11 11.277 & +17 46 31.94 & 0.505 & $1150^{+260 }_{-340}$ & $7.8\pm{0.3}$ & $8.8\pm{0.7}$&  4 & F555W,F814W\\
  J1149.5+2223 & 11 49 35.093 & +22 24 10.94 & 0.544 & $1840^{+120 }_{-170}$ & $17.6\pm{0.4}$ & $9.1\pm{0.7}$ & 4 & F555W,F814W\\
  J1423.8+2404 & 14 23 47.663 & +24 04 40.14 & 0.543 & $1300^{+120}_{-170}$ & $16.5\pm{0.7}$ & $7.0\pm{0.8}$ & 1 & F555W,F814W\\
  J2129.4-0741 & 21 29 26.214 & -07 41 26.22 & 0.589 & $1400^{+170}_{-200}$ & $15.7\pm{0.4}$ & $8.1\pm{0.7}$ & 3 & F555W,F814W\\
  J2214.9-1359 & 22 14 57.415 & -14 00 10.78 & 0.503 & $1300^{+90}_{-100}$ & $14.1\pm{0.3}$ & $8.8\pm{0.7}$ & 2 & F555W,F814W\\
  \hline
\end{tabular}
\end{center}
\end{table*}
\section{Strong Lensing Modelling and Analysis}\label{model}

We apply our well tested approach to lens modelling, which we have
 applied successfully before to various clusters uncovering unprecedentedly
 large numbers of multiply-lensed images in A1689, Cl0024, MACS J1149.5+2223 and MACS J0717.5+3745 (respectively, Broadhurst et al. 2005, Zitrin
 et al. 2009b, Zitrin \& Broadhurst 2009, Zitrin
 et al. 2009a). The full details of this approach can be found in these
 earlier papers. Briefly, the basic assumption adopted is that mass
 approximately traces light, so that the photometry of the red cluster
 member galaxies is the starting point for our model. In a recent paper (Zitrin et al. 2010) we analyse Abell 1703 and show that this assumption is well based, by comparison to an assumption-free non-parametric technique (Liesenborgs et al. 2006) which yields a very similar overall mass distribution. In addition, we show that similar to other clusters we have analysed to date, our modelling method has the inherent flexibility to find and reproduce many multiple-images even if initially constrained by only a few obvious systems.

 As mentioned, the starting point of the model are cluster member galaxies, which are identified as lying close to the cluster sequence by the photometry provided in the Hubble Legacy Archive (in this process sometimes massive foreground galaxies are included as well, scaled down by the relative distance ratio, since they can locally affect nearby images). We then approximate the large scale distribution of matter by assigning a
 power-law mass profile to each galaxy, the sum of which is then
 smoothed. The degree of smoothing ($S$) and the index of the power-law ($q$) are
 the most important free parameters determining the mass profile. A worthwhile improvement in
 fitting the location of the lensed images is generally found by
 expanding to first order the gravitational potential of the smooth
 component, equivalent to a coherent shear describing the overall matter ellipticity, where the direction of the
 shear and its amplitude are free, allowing for some flexibility in
 the relation between the distribution of DM and the distribution of
 galaxies which cannot be expected to trace each other in detail. This freedom also allows the effective centre to slightly shift, as was the case in our analysis of Cl0024 (Zitrin et al. 2009b). The
 total deflection field $\vec\alpha_T(\vec\theta)$, consists of the
 galaxy component, $\vec{\alpha}_{gal}(\vec\theta)$, scaled by a
 factor $K_{gal}$, the cluster DM component
 $\vec\alpha_{DM}(\vec\theta)$, scaled by (1-$K_{gal}$), and the
 external shear component $\vec\alpha_{ex}(\vec\theta)$:

\begin{equation}
\label{defTotAdd}
\vec\alpha_T(\vec\theta)= K_{gal} \vec{\alpha}_{gal}(\vec\theta)
+(1-K_{gal}) \vec\alpha_{DM}(\vec\theta)
+\vec\alpha_{ex}(\vec\theta),
\end{equation}
where the deflection field at position $\vec\theta_m$
due to the external shear,
$\vec{\alpha}_{ex}(\vec\theta_m)=(\alpha_{ex,x},\alpha_{ex,y})$,
is given by:
\begin{equation}
\label{shearsx}
\alpha_{ex,x}(\vec\theta_m)
= |\gamma| \cos(2\phi_{\gamma})\Delta x_m
+ |\gamma| \sin(2\phi_{\gamma})\Delta y_m,
\end{equation}
\begin{equation}
\label{shearsy}
\alpha_{ex,y}(\vec\theta_m)
= |\gamma| \sin(2\phi_{\gamma})\Delta x_m -
  |\gamma| \cos(2\phi_{\gamma})\Delta y_m,
\end{equation}
and $(\Delta x_m,\Delta y_m)$ is the displacement vector of the
position $\vec\theta_m$ with respect to a fiducial reference position,
which we take as the lower-left pixel position $(1,1)$, and
$\phi_{\gamma}$ is the position angle of the spin-2 external
gravitational shear measured anti-clockwise from the $x$-axis.
The normalisation of the model and the relative scaling of the
smooth DM component versus the galaxy contribution brings the
total number of free parameters in the model to 6.

%Note, since
Note that since
our goal is to find the Einstein radius distribution and the masses within the critical curves, and since many clusters lack multiple-images redshift information, we do not attempt to accurately constrain here the mass profiles of the sample clusters and thus the parameters $q$ and $S$ are relatively irrelevant. Any reasonable values for these parameters yield similar critical curves, Einstein radii and
thus also the corresponding masses enclosed within them.
This effectively leaves us with 4 parameters per model which can be easily constrained by the multiple-images incorporated and found here. Still, we have explored also the $q$ and $S$ parameter space for each cluster to verify this. In addition, we showed that in cases where there are too few bands in order to obtain reliable photometric redshifts and where no spectroscopic redshifts are available, a very efficient assumption is that the outer, blue multiple-images are at a redshift of $z\sim2-2.5$. This assumption is based on previous analysis of many clusters for which the vast majority of the outer blue images are at this redshift range due to the \emph{nesting} effect by which the critical curve expands for higher redshift objects, yet the redshift of these blue images is limited by the Lyman-alpha break to be lower than $z\sim3$. For example, this assumption was proven to be very accurate in MACS J1149.5+2223, where the intermediate-distance blue images we assumed at $z\sim2$ (system 3 in Zitrin \& Broadhurst 2009) were later spectroscopically found to be at $z=1.89$, and our corresponding estimation of the spiral galaxy at $z\simeq1.5$ (system 1 in Zitrin \& Broadhurst 2009),  was later verified spectroscopically to be at $z=1.49$ (Smith et al. 2009). In addition, in our works on A1689 and Cl0024 we showed that the same minimum is obtained both when minimising according to the images location in the image-plane, and when minimising according to the photometric redshifts. This means that in practice also the slope which generally can be constrained only using redshifts information, can be well approximated by minimising solely the image-plane RMS of the reproduced images, especially if a clear minimum is seen (see Zitrin et al. 2009b). Note, here we aim to constrain the critical curves and the mass enclosed within them which are both relatively independent and indifferent to the profile, enabling an accurate measurement also without redshift information, which in turn can be used to approximate the magnification of the lens, since this quantity is profile dependent.

Firstly we use our preliminary models to lens various well detected candidate lensed galaxies back to the
source plane using the derived deflection field, and then relens this
source plane to predict the detailed appearance and location of
additional counter images, which may then be identified in the data by
morphology, internal structure and colour.  We stress that multiple images found this way
must be accurately reproduced by our model and are not simply eyeball
``candidates'' requiring redshift verification. Note, due to the volume of this work we do not attempt to exhaust each cluster field finding great numbers of multiple-images, and concentrate instead on finding clear examples of multiply-lensed systems to constrain the models and obtain the Einstein radius. The fit is assessed by the RMS uncertainty in the image plane:

\begin{equation} \label{RMS}
RMS_{images}^{2}=\sum_{i} ((x_{i}^{'}-x_{i})^2 + (y_{i}^{'}-y_{i})^2) ~/ ~N_{ima
ges},
\end{equation}
where $x_{i}^{'}$ and $y_{i}^{'}$ are the locations given by the
model, and $x_{i}$ and $y_{i}$ are the real images location, and the
sum is over all $N_{images}$ images. The best-fit solution is obtained by
the minimum RMS, and the uncertainties are determined by the
location of predicted images in the image plane.

Importantly, this image-plane minimisation does not suffer from the
well known bias involved with source plane minimisation, where
solutions are biased by minimal scatter towards shallow mass profiles
with correspondingly higher magnification. The model is successively
refined as additional sets of multiple images are identified and then
incorporated to improve the fit, using also their redshifts measurements or estimates for better constraining the mass slope through the cosmological relation of the $D_{ls}/D_{s}$ growth (see Zitrin et al. 2009b). We detail this procedure for each cluster in the corresponding subsection below, each followed by corresponding figures of the multiple-images and the critical curves, and the mass distribution (Figures \ref{curves0018} to \ref{contours2214}).

\subsection{MACS J0018.5+1626}
\begin{figure}
 \begin{center}
  \includegraphics[width=80mm,trim=0mm 0mm 0mm 0mm,clip]{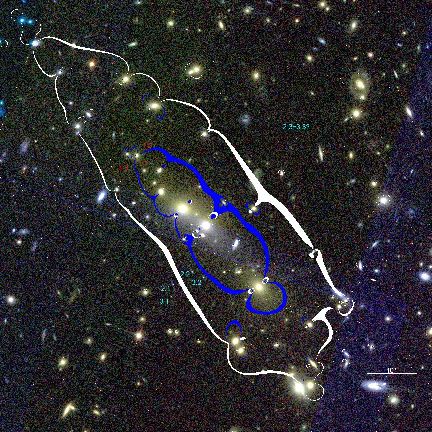}
 \end{center}
\caption{Galaxy cluster MACS J0018.5+1626 ($z=0.546$, also known as Cl 0016+1609, or MS 0016) imaged
with Hubble/ACS F606W, F775W, and F850LP bands. The blue curve
overlaid shows the tangential critical curve corresponding to the
distance of system~1 at an estimated redshift of $z\sim1.5$, and which
passes through the close triplet of lensed arcs in this system. The larger
critical curve overlaid in white corresponds to the larger source
distance estimated as $z\sim2-2.5$, passing through the close pairs of the candidate images of systems 2 and 3. This critical curve encloses a
large lensed region, with an equivalent Einstein radius of $\sim150~kpc$ at the
redshift of the cluster. Note, our model predicts tiny images of pairs 2.1/2.2 and 3.1/3.2 at the other side of the cluster, which we are not able to securely detect due to poor depth and lack of colour range.}
\label{curves0018}
\end{figure}

\begin{figure}
 \begin{center}
  \includegraphics[width=80mm, trim=5mm 0mm 5mm 5mm,clip]{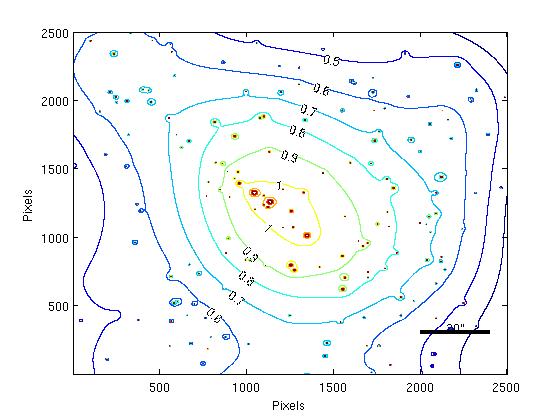}
 \end{center}
\caption{2D surface mass distribution ($\kappa$), in units of the
critical density, of MACS J0018.5+1626. Contours are shown in linear units, derived from
the mass model constrained using the multiply-lensed images seen in
Figure \ref{curves0018}. Axes are in ACS pixels ($0.05 \arcsec /pixel$), and a $20\arcsec$ scale bar is overplotted.}
\label{contours0018}
\end{figure}

The galaxy cluster MACS J0018.5+1626 ($z=0.546$, also known as Cl 0016+1609, or MS 0016) has been subject to extensive study, mainly due to its high X-ray luminosity, strong SZ effect, and high redshift (e.g., Dressler \& Gunn 1992, Dressler et al. 1999, Luppino et al. 1999, Ellis \& Jones 2002, Clowe et al. 2000, 2003, Laroque et al. 2003), setting an example of a rich distant cluster (Worrall \& Birkinshaw, 2003, and references therein). It has been established that this cluster is the major part of a large-scale filamentary structure (Worrall \& Birkinshaw, 2003, and references therein, Tanaka, 2007). This fact and the prominent X-ray emission imply a high cluster mass, which should be manifested in multiply-lensed systems spread throughout the field. However, due to lower quality optical data and the complexity of its structure, no multiply-lensed systems were found in MACS J0018.5+1626 up to date. Only one strongly-lensed arc was referred to before in this cluster (Lavery 1996), yet no spectroscopic redshift nor counter-images were introduced.
We begin our analysis by constructing an initial model in the method described above (\S \ref{model}), where as a starting point we insert averaged values to the various parameters, since no system is known a-priori to constrain the fit. We then notice the similar colours and symmetry of the images of systems 1, 2, and 3. After verifying that the reproduction of these systems is physically likely in the context of our model, we use these systems to fully constrain the model and derive the critical curves (see Figure \ref{curves0018}), their corresponding Einstein radius ($\theta_{e}=28\pm2 \arcsec$ for a source at $z_{s}\sim2$; systems 2 \& 3) and the mass enclosed within them ($1.46 \pm 0.1 \times 10^{14} M_{\odot}$), see Figures \ref{curves0018} and \ref{contours0018} for more information. Accordingly, the estimated redshift of system 1 is $z_{s}\sim1.5$.

There are various other mass estimates
for this cluster, mainly from weak-lensing analyses. Smail et al. (1997) quote values of $\sim 2.9 \times 10^{14} M_{\odot}$ and $\sim 1.87 \times 10^{14} M_{\odot}$ within 200 kpc (depending on the method, see also Smail et al. 1995), and Hoekstra (2007) quotes a value of $\sim 7.9 \times 10^{14} M_{\odot}$ within 500 kpc, all in general accordance with our measurement. The Einstein radius for this cluster was also mentioned before. Hoekstra (2007) quotes $\sim 9 \arcsec$ from a SIS model fit to the tangential
distortion from 0.25 to 1.5 $h^{-1}$ Mpc, while Williams, Navarro \& Bartelmann (1999) quote $25 \arcsec$ according to the arc discussed by Lavery (1996), in accordance with our estimation.
The reference centre of our analysis is fixed near the centre of the F775W (program ID 10493) ACS frame at: RA = 00:18:33.41, Dec = +16:26:14.95 (J2000.0), where one arcsecond corresponds to 6.42 kpc at the redshift of this cluster.

\subsection{MACS J0025.4-1222}
\begin{figure}
 \begin{center}
  \includegraphics[width=90mm,trim=0mm 0mm 0mm 0mm,clip]{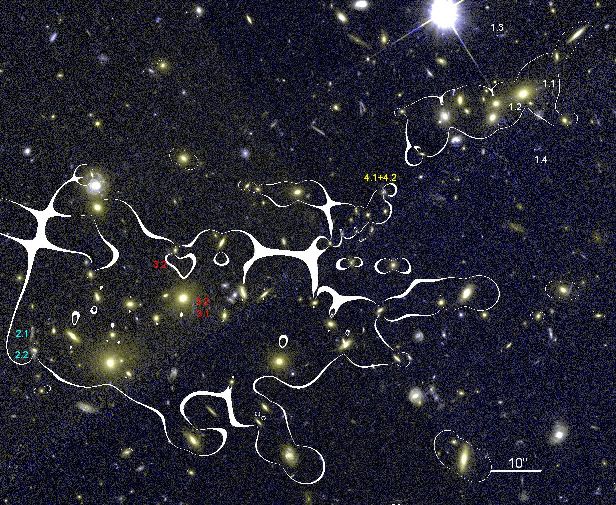}
 \end{center}
\caption{Galaxy cluster MACS J0025.4-1222 ($z=0.584$) imaged
with Hubble/ACS F555W and F814W bands. The overlaid critical curve (white) corresponds to system 1 ($z_{s}=2.38$), enclosing a critical area of an effective Einstein radius of $\sim 200$ kpc at the redshift of this cluster.}
\label{curves0025}
\end{figure}

\begin{figure}
 \begin{center}
 \includegraphics[width=80mm,trim=0mm 0mm 0mm 0mm,clip]{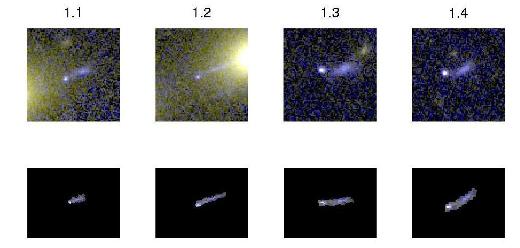}
 \end{center}
\caption{Reproduction of system 1 in MACS J0025.4-1222 by our model, by delensing image 1.1 into the source plane, and relensing the source-plane pixels onto the image plane to accurately form the other images.}
\label{rep0025}
\end{figure}

The galaxy cluster MACS J0025.4-1222 ($z=0.584$) and its merging properties were recently analysed by Brada\v{c} et al. (2008b), which found multiply-lensed images of 4 background galaxies, and obtained spectroscopic redshifts for a pair of them and
photometric redshifts for the other two systems: system A+B
(system 1 here) at $z_{spec}=2.38$, system C (system 2 here) at
$z_{phot}=1^{+0.5}_{-0.2}$, and system D (system 3 here) at
$z_{phot}=2.8^{+0.4}_{-1.8}$. We use systems 1 and 2 (see Figure \ref{curves0025}) and one more
uncovered arc (system 4 in Figure \ref{curves0025}; $z_{s}\sim1.9$), to derive the corresponding Einstein radius ($\theta_{e}= 30\pm2 \arcsec$ for $z_{s}\sim2.38$) and the mass enclosed within it ($2.42^{+0.10}_{-0.13}\times 10^{14} M_{\odot}$), in excellent agreement with the result of Brada\v{c} et al (2008b; $\simeq 2.5 \times 10^{14} M_{\odot}$, within 300 kpc centred on the SE BCG). In addition we find that system A in Brada\v{c} et al (2008b; system 1 here) includes also an additional arc on the other side of the BCG, as seen in Figure \ref{curves0025} (arc 1.2), which our model reproduces very accurately (see Figure \ref{rep0025}). The reference centre of our analysis is fixed near the centre of the ACS frame at: RA = 00:25:29.61, Dec = -12:22:52.89 (J2000.0), where one arcsecond corresponds to 6.64 kpc at the redshift of this cluster.

\begin{figure}
 \begin{center}
  \includegraphics[width=80mm, trim=5mm 0mm 5mm 5mm,clip]{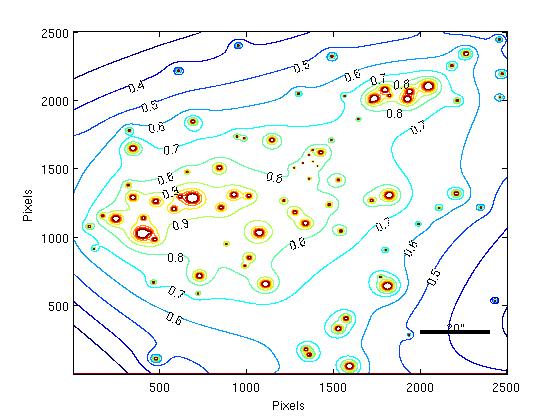}
 \end{center}
\caption{2D surface mass distribution ($\kappa$), in units of the
critical density, of MACS J0025.4-1222. Contours are shown in linear units, derived from
the mass model constrained using the multiply-lensed images seen in
Figure \ref{curves0025}. Axes are in ACS pixels ($0.05 \arcsec /pixel$), and a $20\arcsec$ scale bar is overplotted.}
\label{contours0025}
\end{figure}
\subsection{MACS J0257.1-2325}
\begin{figure}
 \begin{center}
  \includegraphics[width=80mm,trim=0mm 0mm 0mm 0mm,clip]{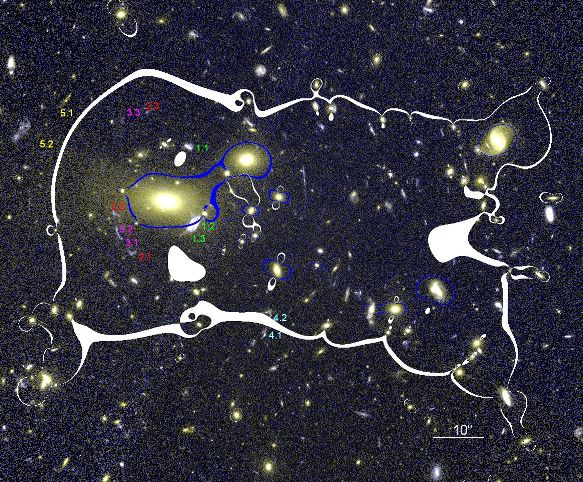}
 \end{center}
\caption{Galaxy cluster MACS J0257.1-2325 ($z=0.505$) imaged
with Hubble/ACS F555W and F814W bands. The tangential critical curves overlaid in blue correspond to a source redshift of $z_{s}\sim1$, passing through the close pair of images 1.2 and 1.3. The larger critical curves overlaid in white correspond to a source redshift of $z_{s}\sim2-2.5$, passing through the close pair of images 4.1 and 4.2.}
\label{curves0257}
\end{figure}

\begin{figure}
 \begin{center}
 \includegraphics[width=60mm,trim=0mm 0mm 0mm 0mm,clip]{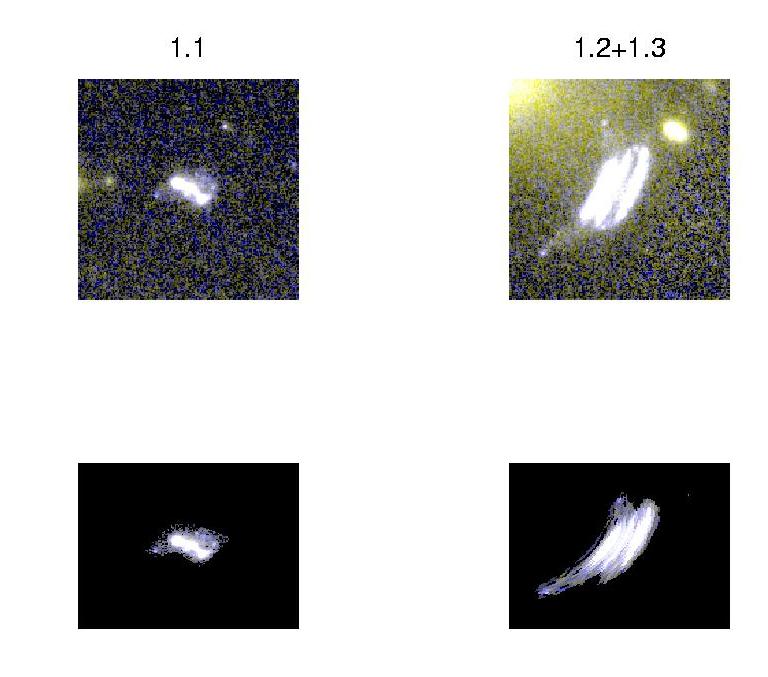}
 \end{center}
\caption{Reproduction of system 1 in MACS J0257.1-2325 by our model, by delensing image 1.1 into the source plane, and relensing the source-plane pixels onto the image plane to accurately form the other images. Note, our model predicts an extra tiny image near the core of the cD covered by its light.}
\label{rep0257}
\end{figure}

In the galaxy cluster MACS J0257.1-2325 ($z=0.505$) various arcs are seen, from which we use 13 multiply-lensed images, belonging to $\sim7$ background distant galaxies, in the redshift range $z\sim1$ to $z\sim4$ to fully constrain the model. The reference centre of our analysis is fixed near the centre of the ACS frame at: RA = 02:57:07.24, Dec = -23:26:02.88 (J2000.0), where one arcsecond corresponds to 6.17 kpc at the redshift of this cluster.
This cluster was not analysed before and we did not find any arc redshift information. We assume a redshift of  $z_{s}\sim2-2.5$ for the faint distant blue images (system 4 in Figure \ref{curves0257}) according to which the spectacular images next to the cD galaxy (system 1; see also Figure \ref{rep0257}) are at $z_{s}\sim1$, the blue images of systems 2 and 3 are at $z_{s}\sim1.5-2$, and the red drop-out candidates are at $z_{s}\sim3.5-4$ (system 5) helping to constrain the fit. We find that the critical curves (Figure \ref{curves0257}) have an equivalent Einstein radius of $39\pm2 \arcsec$ (for $z_{s}\sim2$) and enclose a mass of $3.35^{+0.58}_{-0.10} \times 10^{14} M_{\odot}$.
.

\begin{figure}
 \begin{center}
  \includegraphics[width=85mm,trim=5mm 0mm 5mm 5mm,clip]{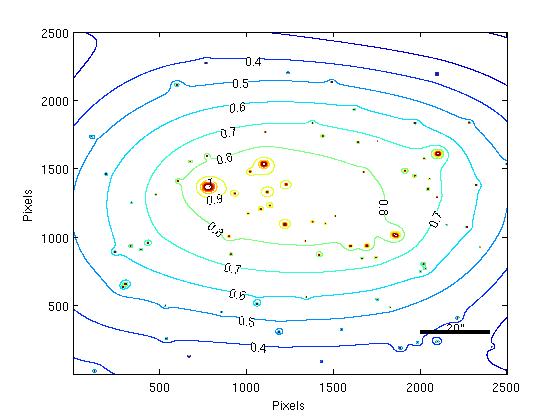}
 \end{center}
\caption{2D surface mass distribution ($\kappa$), in units of the
critical density, of MACS J0257.1-2325. Contours are shown in linear units, derived from
the mass model constrained using the multiply-lensed images seen in
Figure \ref{curves0257}. Axes are in ACS pixels ($0.05 \arcsec /pixel$), and a $20\arcsec$ scale bar is overplotted.}
\label{contours0257}
\end{figure}

\subsection{MACS J0454.1-0300}
\begin{figure}
 \begin{center}
  \includegraphics[width=85mm,trim=0mm 0mm 0mm 0mm,clip]{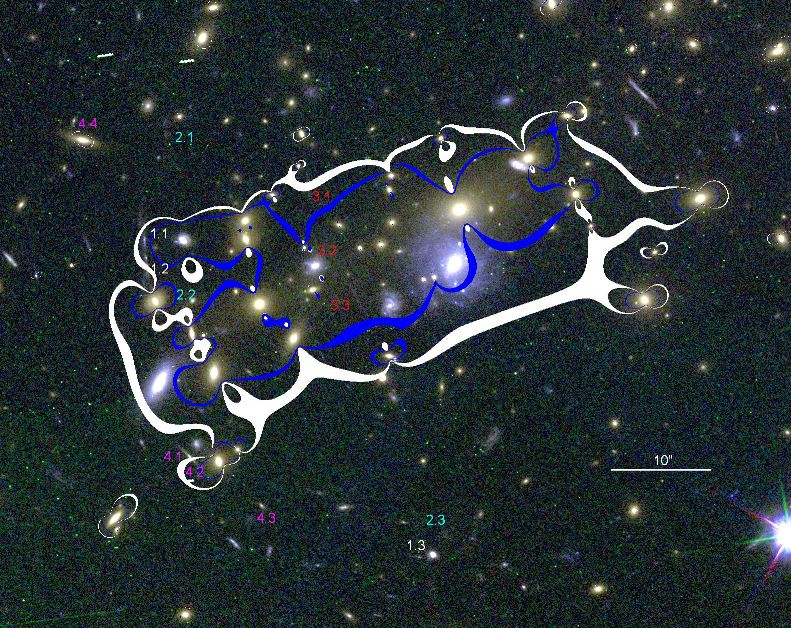}
 \end{center}
\caption{Galaxy cluster MACS J0454.1-0300 ($z=0.538$, also known as MS 0451.6-0305) imaged
with Hubble/ACS F555W, F775W, and F814W bands. The tangential critical curves overlaid in white correspond to a source redshift of $z_{s}=2.9$ (system 1, see also Borys et al. 2004), passing through the close pair of images 1.1/1.2, and 4.1/4.2. The smaller critical curves overlaid in blue correspond to a source redshift of $z_{s}\sim2$.}
\label{curves0454}
\end{figure}

The galaxy cluster MACS J0454.1-0300 ($z=0.538$, also known as MS 0451.6-0305) has been subject to extensive study due to its high X-ray luminosity, strong SZ effect, and high redshift (e.g., Ellingson et al. 1998, Molnar et al. 2002, Laroque et al. 2003, Geach et al. 2006), and is known to host various lensed sub-mm sources with radio counterparts (Takata et al. 2003, Borys et al. 2004, Berciano Alba et al. 2007, 2009), which we incorporate in our analysis below.
Previous strong-lensing models of this cluster were introduced by Borys et al. (2004) and Berciano Alba et al. (2009) with similar symmetry as our model.
We fully constrain our model with the arcs and their redshift information listed in these papers (see also Figure \ref{curves0454} here), from which we derive an effective Einstein radius of $19\pm2 \arcsec$ for a source redshift of $z_{s}=2.9$ (system 1 in Figure \ref{curves0454}) and a mass of $0.82^{+0.03}_{-0.01} \times 10^{14} M_{\odot}$ enclosed within this critical curve. For a source redshift of $z_{s}\sim2$ we correspondingly get an effective Einstein radius of $13\pm2 \arcsec$, and a mass of $0.41^{+0.03}_{-0.01} \times 10^{14} M_{\odot}$ enclosed within the critical curve (see Figure \ref{curves0454}). Berciano Alba et al. (2009) quote a mass of $1.73 \times 10^{14} M_{\odot}$ within 30 $\arcsec$ of the cluster centre. Our model yields $1.8 \times 10^{14} M_{\odot}$ within 30 $\arcsec$ of the cluster centre, in full agreement. In our modelling process we find two other systems of multiply-lensed images. The first (system 3) is a triplet of faint arcs next to the core, at an estimated redshift of $z_{s}\sim1.5-2$. The second system (number 4) consists of 4 bright images, all have prominent emission in the K'-band (see Takata et al. 2003), corresponding to a redshift of $z_{s}\simeq 2.9$ similar to system 1. See Figure \ref{curves0454} for more details.
The reference centre of our analysis is fixed near the centre of the F555W (proposal ID 9722) ACS frame at: RA = 04:54:10.36, Dec = -03:01:03.02 (J2000.0), where one arcsecond corresponds to 6.37 kpc at the redshift of this cluster.

\begin{figure}
 \begin{center}
  \includegraphics[width=80mm,trim=5mm 0mm 5mm 5mm,clip]{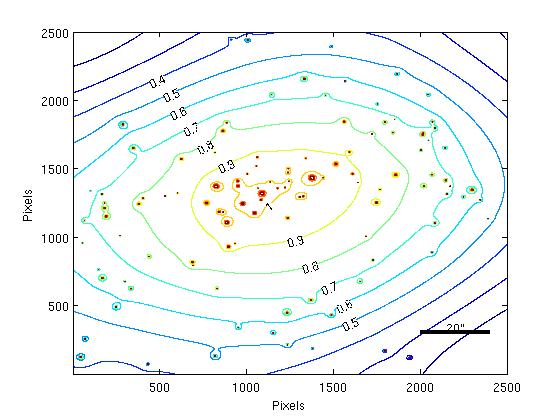}
 \end{center}
\caption{2D surface mass distribution ($\kappa$), in units of the
critical density, of MACS J0454.1-0300. Contours are shown in linear units, derived from
the mass model constrained using the multiply-lensed images seen in
Figure \ref{curves0454}. Axes are in ACS pixels ($0.05 \arcsec /pixel$), and a $20\arcsec$ scale bar is overplotted.}
\label{contours0454}
\end{figure}
\subsection{MACS J0647.7+7015}
\begin{figure}
 \begin{center}
  \includegraphics[width=80mm,trim=0mm 0mm 0mm 0mm,clip]{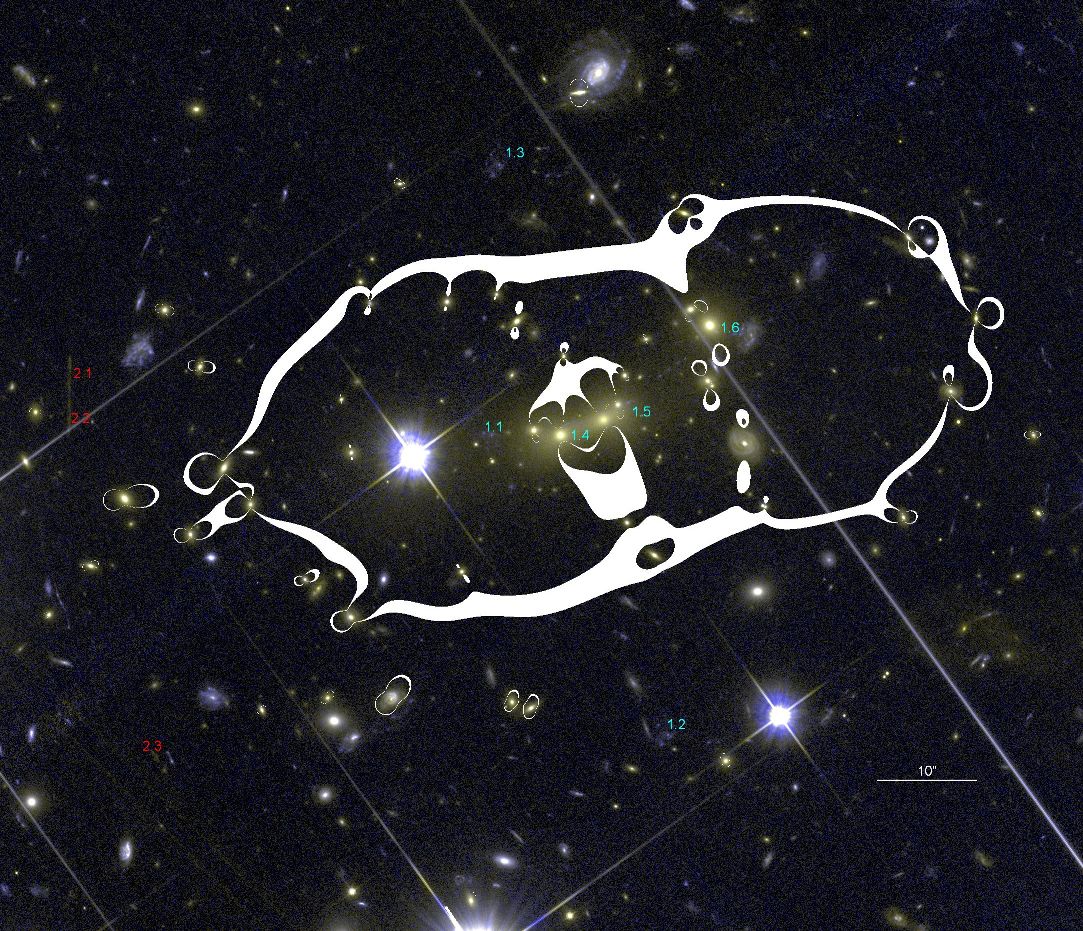}
 \end{center}
\caption{Galaxy cluster MACS J0647.7+7015 ($z=0.591$) imaged
with Hubble/ACS F555W and F814W bands. The critical curves overlaid in white correspond to the six images we identified as belonging to system 1, at an estimated source redshift of $z_{s}\sim2$, enclosing a critical area with an effective Einstein radius of $\sim 190$ kpc, at the redshift of this cluster. The second system consists of 3 drop-out candidate images, corresponding to a redshift of $z_{s}\sim3.5$.}
\label{curves0647}
\end{figure}

In the galaxy cluster MACS J0647.7+7015 ($z=0.591$) we note a remarkable blue background system lensed 6 times, where each lensed image is a quartet of arcs, spread across the image and between the BCGs. Additional blue faint images between images 1.5 and 1.6 (see Figure \ref{curves0647}) may also be related to this system. We did not find any record of past analysis nor redshift information and we assume the redshift of this system to be $z_{s}\sim2$ corresponding to an equivalent Einstein radius of $28\pm2 \arcsec$ enclosing a mass of $2.07\pm0.10 \times 10^{14} M_{\odot}$. This also corresponds to the higher redshift assumed for system 2, which is estimated to be a drop-out galaxy at $z_{s}\sim3.5$ helping us to constrain the fit. See Figure \ref{curves0647} for more details. The reference centre of our analysis is fixed between the two cD galaxies at: RA = 06:47:50.23, Dec = +70:14:55.37 (J2000.0), where one arcsecond corresponds to 6.67 kpc at the redshift of this cluster.

\begin{figure}
 \begin{center}
  \includegraphics[width=80mm,trim=5mm 0mm 5mm 5mm,clip]{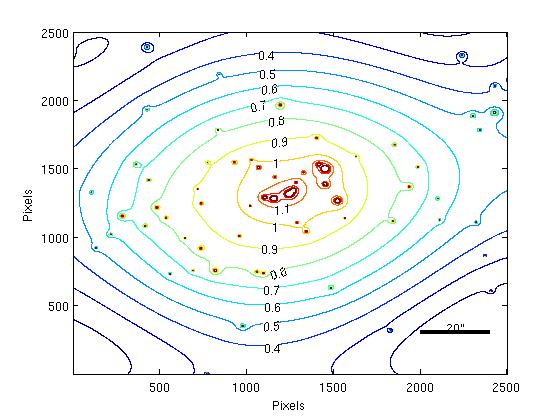}
 \end{center}
\caption{2D surface mass distribution ($\kappa$), in units of the
critical density, of MACS J0647.7+7015. Contours are shown in linear units, derived from
the mass model constrained using the multiply-lensed images seen in
Figure \ref{curves0647}. Axes are in ACS pixels ($0.05 \arcsec /pixel$), and a $20\arcsec$ scale bar is overplotted.}
\label{contours0647}
\end{figure}

\subsection{MACS J0717.5+3745}
\begin{figure}
 \begin{center}
  \includegraphics[width=80mm,trim=0mm 0mm 0mm 0mm,clip]{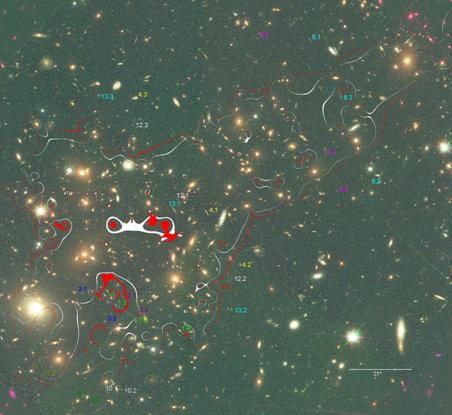}
 \end{center}
\caption{Galaxy cluster MACS J0717.5+3745 ($z=0.546$) imaged
with Hubble/ACS F555W, F606W, and F814W bands. The 34 multiply lensed
images identified by our model are numbered here. The white curve
overlaid shows the tangential critical curve corresponding to the
distance of system~1 at an estimated redshift of $z\sim2-2.5$, and which
passes through several close pairs of lensed images in this system. The larger
critical curve overlaid in red corresponds to the larger source
distance for the red dropout galaxy number ~5, at the estimated photometric
redshift of $z\sim3.5-4$. This large tangential critical curve encloses a
very large lensed region equivalent to $\sim400~kpc$ in radius at the
redshift of the cluster. Figure was originally published in Zitrin et al. 2009a.}
\label{curves0717}
\end{figure}

The galaxy cluster MACS J0717.5+3745 ($z=0.55$) was analysed recently by Zitrin et al. (2009a). This very X-ray luminous cluster is  the denser north-western region of the large-scale filament found by Ebeling et al. (2004) and it is the largest known lens, with an equivalent Einstein radius of 55 \arcsec and a mass of $7.4\pm0.5  \times 10^{14} M_{\odot}$. 13 multiply-lensed systems were used to constrain the fit as seen in Figure \ref{curves0717} taken from Zitrin et al. (2009a). Due to its very large critical area, this cluster is a great target for finding high-$z$ objects, and has been recently proposed for uncovering early stars (or ``dark stars'', Zackrisson et al. 2010).

\begin{figure}
 \begin{center}
  \includegraphics[width=80mm,trim=5mm 0mm 5mm 5mm,clip]{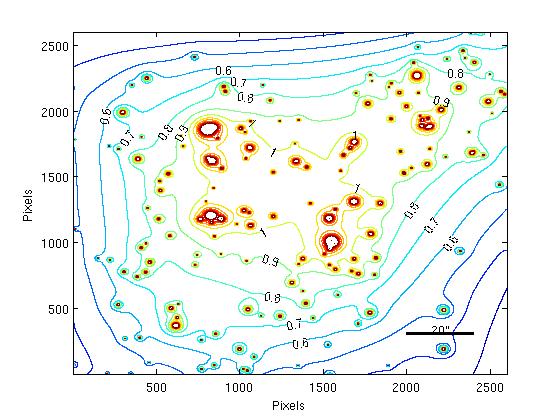}
 \end{center}
\caption{2D surface mass distribution ($\kappa$), in units of the
critical density, of the inner central region of MACS J0717.5+3745. Contours are shown in linear units, derived from
the mass model constrained using 34 multiply-lensed images seen in
Figure \ref{curves0717}. Axes are in ACS pixels ($0.05 \arcsec /pixel$), and a $20\arcsec$ scale bar is overplotted. Note that the central mass distribution of
is rather flat reflecting the unrelaxed appearance
of this cluster. Figure was originally published in Zitrin et al. 2009a.}
\label{contours0717}
\end{figure}

\subsection{MACS J0744.8+3927}
\begin{figure}
 \begin{center}
  \includegraphics[width=85mm,trim=0mm 0mm 0mm 0mm,clip]{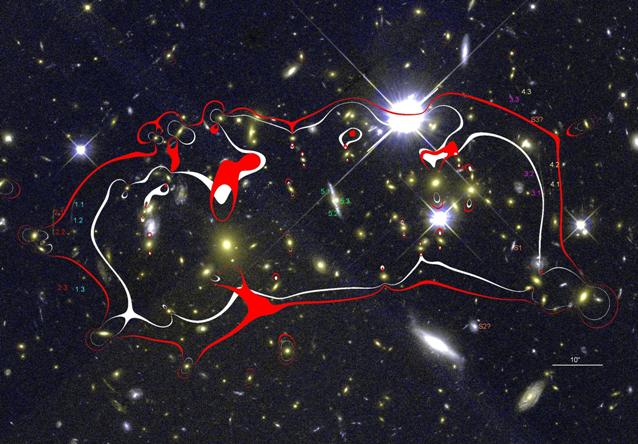}
 \end{center}
\caption{Galaxy cluster MACS J0744.8+3927 ($z=0.698$) imaged
with Hubble/ACS F555W and F814W bands. The critical curves overlaid in red correspond to systems 2 and 4, at an estimated source redshift of $z_{s}\sim3.5$. The inner critical curve overlaid in white corresponds to a source redshift of $z_{s}\simeq 2.2$ (arc ``S1''; measured spectroscopically to be at this redshift by Jones et al. 2009). We mark possible counter images of this arc and other multiply-lensed images we uncovered throughout the frame.}
\label{curves0744}
\end{figure}

The luminous X-ray cluster MACS J0744.8+3927 ($z=0.698$) is the highest-$z$ cluster of this sample, comprising several obvious close pairs of
multiply-lensed galaxies which are immediately visible throughout the frame, with which we begin our modelling process (see Figure \ref{curves0744}). Systems 2 and 4 are likely drop-outs and we estimate their redshift as $z_{s}\sim3.5$ which helps to constrain our model. These systems correspond to similar lensing distance ratios thus basing this assumption. We did not find record of past analysis yet recently a spectroscopic redshift of a resolved galaxy behind this cluster was published (Jones et al. 2009; $z_{s}=2.21$, marked as ``S1'' in Figure \ref{curves0744} here). Unfortunately, no counter images were reported. As can be seen in Figure \ref{curves0744}, this arc lies within the critical curve for a source at $z_{s}\simeq 2.2$ according to our model, and therefore should be multiply-lensed and we mark other possible counter images of this galaxy. Another option, which would favour a steeper model, is that the three clumps seen in this arc are counter images of the same galaxy, though we find this option to be less probable according to the velocity map presented in Jones et al. (2009), which shows that the three clumps have different velocities. In addition, another less probable case is that the critical curve for a source at $z_{s}\simeq 2.2$ is much smaller, though this does not agree with the other assumed redshifts for the outer blue arcs (system 1 for example), which are not likely be at a higher redshift than $z_{s}\sim3$ since they are closer to the cluster center than system 2 but following the same symmetry, yet clearly seen in the F555W band. Deeper imaging in more passbands will help to uniquely solve the strong-lensing in this cluster. We derive an equivalent Einstein radius of $31\pm2 \arcsec$ for a source redshift of $z_{s}\simeq 2.2$, enclosing a mass of $3.1\pm0.1 \times 10^{14} M_{\odot}$. The reference centre of our analysis is fixed on the cD galaxy at: RA = 07:44:52.80, Dec = +39:27:26.50 (J2000.0). One arcsecond corresponds to 7.17 kpc at the redshift of this cluster.

\begin{figure}
 \begin{center}
  \includegraphics[width=80mm,trim=5mm 0mm 5mm 5mm,clip]{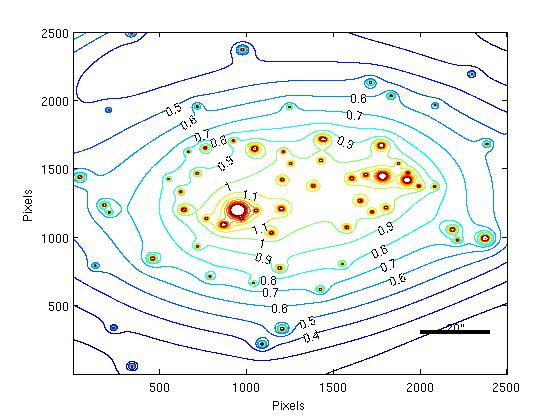}
 \end{center}
\caption{2D surface mass distribution ($\kappa$), in units of the
critical density, of MACS J0744.8+3927. Contours are shown in linear units, derived from
the mass model constrained using the multiply-lensed images seen in
Figure \ref{curves0744}. Axes are in ACS pixels ($0.05 \arcsec /pixel$), and a $20\arcsec$ scale bar is overplotted.}
\label{contours0744}
\end{figure}

\subsection{MACS J0911.2+1746}
\begin{figure}
 \begin{center}
  \includegraphics[width=80mm,trim=0mm 0mm 0mm 0mm,clip]{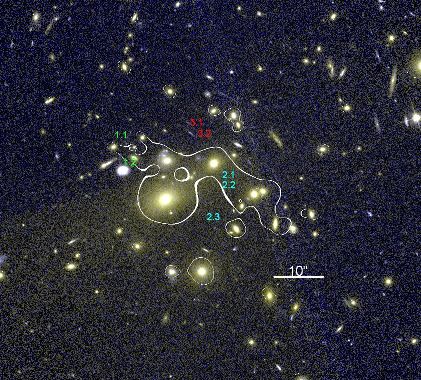}
 \end{center}
\caption{Galaxy cluster MACS J0911.2+1746 ($z=0.505$) imaged
with Hubble/ACS F555W and F814W bands. The 7 multiply-lensed
images identified by our model are numbered here. The white curve
overlaid shows the tangential critical curve corresponding to the
distance of system~1 at an estimated redshift of $z\sim2$. This cluster comprises the smallest Einstein radius and mass of the analysed sample. The area enclosed within the marked critical curve corresponds to an equivalent Einstein radius of 68 kpc at the cluster redshift.}
\label{curves0911}
\end{figure}

\begin{figure}
 \begin{center}
  \includegraphics[width=80mm,trim=5mm 0mm 5mm 5mm,clip]{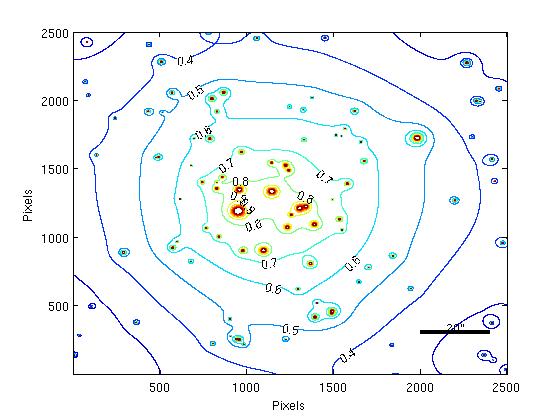}
 \end{center}
\caption{2D surface mass distribution ($\kappa$), in units of the
critical density, of MACS J0911.2+1746. Contours are shown in linear units, derived from
the mass model constrained using the multiply-lensed images seen in
Figure \ref{curves0911}. Axes are in ACS pixels ($0.05 \arcsec /pixel$), and a $20\arcsec$ scale bar is overplotted.}
\label{contours0911}
\end{figure}

In the galaxy cluster MACS J0911.2+1746 ($z=0.505$) not many prominent arcs are seen, in agreement with the small critical area we derived ($\theta_{e}=11^{+3}_{-1} \arcsec$, for a source redshift of $z_{s}\sim2$). We found no record of previous SL analysis of this cluster nor arcs redshift information. Only a few arcs are seen throughout the frame, from which we are able to match 7 multiply-lensed images which are iteratively incorporated into the model, belonging to 3 distant galaxies. Two of these galaxies (systems 1 and 2) are to our estimation at $z_{s}\sim2$. We find that the mass enclosed within the critical curves (for a source redshift of $z_{s}\sim2$) is $0.28^{+0.02}_{-0.01} \times 10^{14} M_{\odot}$. This critical curve and the enclosed mass are both the smallest of the current sample. See Figures \ref{curves0911} and \ref{contours0911} for more details.
The reference centre of our analysis is fixed near the centre of the ACS frame at: RA = 09:11:10.30, Dec = +17:46:39.33 (J2000.0), where one arcsecond corresponds to 6.17 kpc at the redshift of this cluster.

\subsection{MACS J1149.5+2223}
\begin{figure}
 \begin{center}
  \includegraphics[width=80mm,trim=0mm 0mm 0mm 0mm,clip]{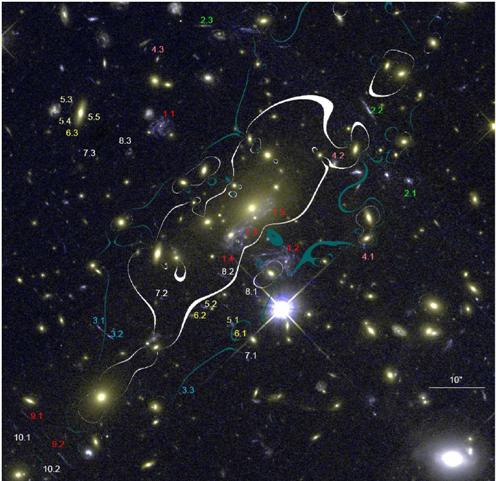}
 \end{center}
\caption{Large scale view of the multiply-lensed galaxies identified by our
model in MACS J1149.5+2223 ($z=0.544$). In addition to the large spiral galaxy system~1, many other
fainter sets of multiply lensed galaxies are uncovered by our
model. The white curve overlaid shows the tangential critical curve
corresponding to the lensing distance of system~1. The larger critical curve
overlaid in blue corresponds to the average distance of the fainter
systems, passing through close pairs of lensed images in systems 2 and
3. This large scale elongated ``Einstein ring'' encloses
a very large critically lensed region equivalent to $170~kpc$ in radius. For this cluster one
arcsecond corresponds to 6.4 kpc, with the standard cosmology. Figure was originally published in Zitrin \& Broadhurst 2009.}
\label{curves11495}
\end{figure}

This cluster was first analysed by us (see Zitrin \& Broadhurst 2009) noting several large blue spiral galaxy
images which are clearly visible near the central brightest cluster galaxy
(Figure \ref{curves11495}). Shortly after, spectroscopic redshifts were published (Smith et al. 2009) in full agreement with the predictions of our analysis: we assumed $z\simeq1.5$ for the spiral-galaxy (system 1), and $z\simeq2$ for the outer blue images (system 3 in Figure \ref{curves11495}), which were later verified to be at $z=1.49$ and $z=1.89$, respectively. Many other faint lensed galaxies are also visible, most of which we have been able to securely identify as belonging to 10 sets of multiply-lensed background galaxies (Figure
%\ref{curves11495}). See
\ref{curves11495}); see
Zitrin \& Broadhurst (2009) for more details. We find that the critical curves for a source redshift of $z\simeq2$ enclose a mass of $1.71\pm0.20 \times 10^{14} M_{\odot}$, and have an equivalent Einstein radius of $27\pm3 \arcsec$.

\begin{figure}
 \begin{center}
  \includegraphics[width=80mm,trim=5mm 0mm 5mm 5mm,clip]{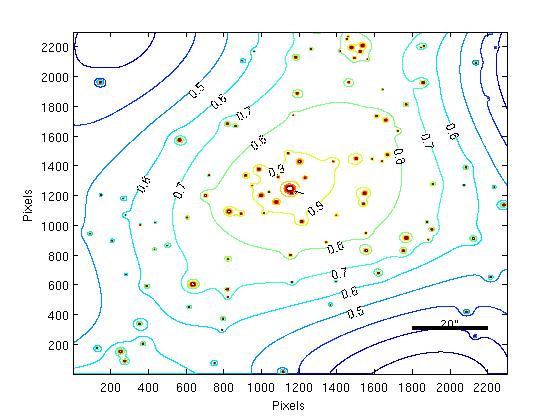}
 \end{center}
\caption{2D surface mass distribution ($\kappa$), in units of the critical density, of MACS J1149.5+2223. Contours are shown in linear units, derived from the mass model constrained using 33 multiply lensed images seen in Figure \ref{curves11495}. Note, the central mass distribution is shallow, and rounder in shape than the critical curves. Figure was originally published in Zitrin \& Broadhurst 2009. Axes are in ACS pixels ($0.05 \arcsec /pixel$), and a $20\arcsec$ scale bar is overplotted.}
\label{contours11495}
\end{figure}

\subsection{MACS J1423.8+2404}
\begin{figure}
 \begin{center}
  \includegraphics[width=80mm,trim=0mm 0mm 0mm 0mm,clip]{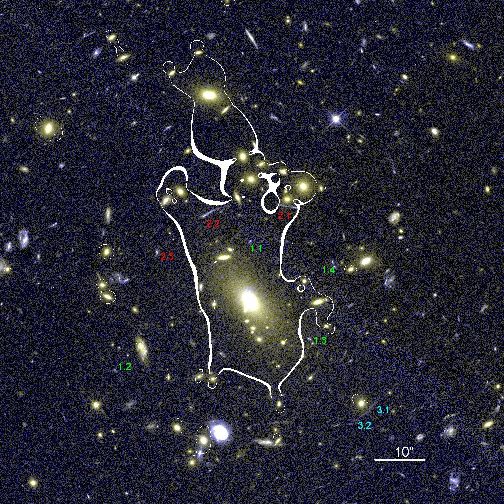}
 \end{center}
\caption{Galaxy cluster MACS J1423.8+2404 ($z=0.543$) imaged
with Hubble/ACS F555W and F814W bands. The multiple-images used in our model are marked on the image. System 1 was spectroscopically measured by Limousin et al. (2009) to be at $z_{s}=2.84$ and system 2 was spectroscopically measured by them to be at $z_{s}=1.78$. On the image we overlay the critical curves for a source redshift of $z_{s}\sim2$, enclosing a critical area with equivalent Einstein radius of $\sim130$ kpc at the cluster redshift.}
\label{curves1423}
\end{figure}

The galaxy cluster MACS J1423.8+2404 ($z=0.543$) was recently analysed by Limousin et al. (2009), finding 3 sets of multiply-lensed galaxies. System 1 here was spectroscopically measured by them to be at $z_{s}=2.84$ and system 2 here was spectroscopically measured by them to be at $z_{s}=1.78$. We incorporate these images in order to fully constrain our model, with an additional locally-lensed arc found here (system 3). The corresponding critical curves for a source redshift of $z_{s}\sim2$ are overlaid in Figure \ref{curves1423}, enclosing a mass of $1.3\pm0.40 \times 10^{14} M_{\odot}$ and yielding an effective Einstein radius of $\theta_{e}=20\pm2 \arcsec$. This Einstein radius is in agreement with that quoted by Limousin et al. (2009), yet the mass is lower than quoted by them, but fully agrees with the Mass/Critical-area relation seen in Figure \ref{reme}, which supports our measurement.
The reference centre of our analysis is fixed near the centre of the ACS frame at: RA = 14:23:48.05, Dec = +24:05:00.23 (J2000.0), where one arcsecond corresponds to 6.40  kpc at the redshift of this cluster.

\begin{figure}
 \begin{center}
  \includegraphics[width=80mm,trim=5mm 0mm 5mm 5mm,clip]{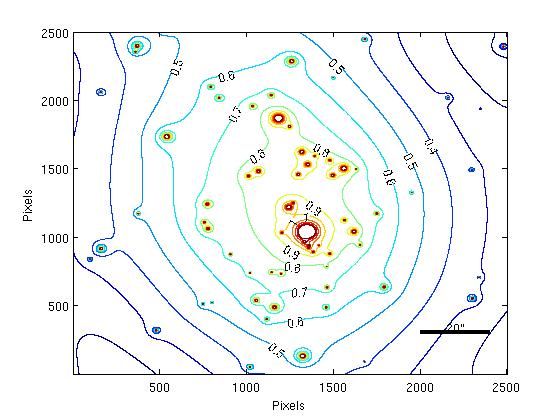}
 \end{center}
\caption{2D surface mass distribution ($\kappa$), in units of the
critical density, of MACS J1423.8+2404. Contours are shown in linear units, derived from
the mass model constrained using the multiply-lensed images seen in
Figure \ref{curves1423}. Axes are in ACS pixels ($0.05 \arcsec /pixel$), and a $20\arcsec$ scale bar is overplotted.}
\label{contours1423}
\end{figure}

\subsection{MACS J2129.4-0741}
\begin{figure}
 \begin{center}
  \includegraphics[width=80mm,trim=0mm 0mm 0mm 0mm,clip]{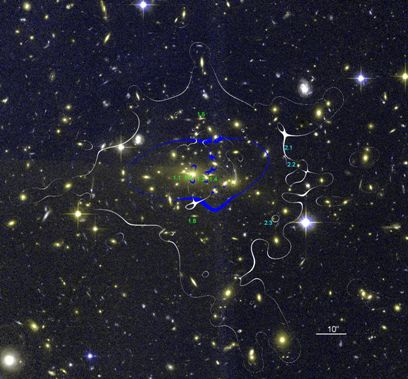}
 \end{center}
\caption{Galaxy cluster MACS J2129.4-0741 ($z=0.589$) imaged
with Hubble/ACS F555W and F814W bands. The large white critical curve corresponds to system 2, at an estimated redshift of $z_{s}\sim2$, enclosing a critical area of an effective Einstein radius of $\sim250$ kpc at the redshift of this cluster. Comprising 6 remarkable images in the centre of the image, system 1 is at a redshift of $z_{s}\sim1$, whose critical curve is overlaid in blue.}
\label{curves2129}
\end{figure}

In the galaxy cluster MACS J2129.4-0741 ($z=0.589$) various arcs are seen throughout the image. One spectacular system consists of 6 lensed images in similar colours as the cluster members (system 1; see Figure \ref{curves2129}). We use this system and the large arc (system 2) further away from the centre to fully constrain the model. Estimating that the outer blue arc (system 2) is at $z_{s}\sim2-2.5$, correspondingly the six-times lensed galaxy of system 1 is at a redshift of $z_{s}\sim1-1.5$. The outer critical curve, corresponding to system 2 encloses an area of an effective radius of $\theta_{e}=37\pm2 \arcsec$ and a mass of $3.4^{+0.6}_{-0.3} \times 10^{14} M_{\odot}$. The reference centre of our analysis is fixed at: RA = 21:29:26.123, Dec = -07:41:27.28 (J2000.0), where one arcsecond corresponds to 6.66 kpc at the redshift of this cluster.
\begin{figure}
 \begin{center}
  \includegraphics[width=80mm,trim=5mm 0mm 5mm 5mm,clip]{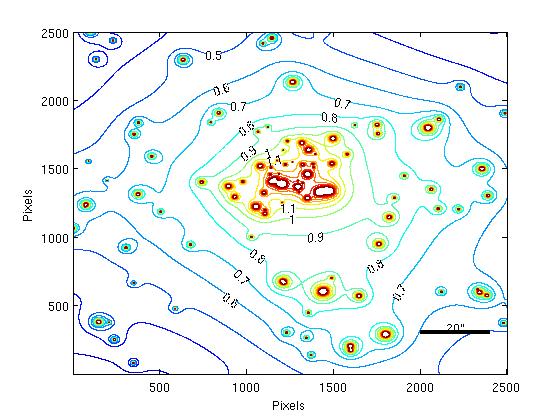}
 \end{center}
\caption{2D surface mass distribution ($\kappa$), in units of the
critical density, of MACS J2129.4-0741. Contours are shown in linear units, derived from
the mass model constrained using the multiply-lensed images seen in
Figure \ref{curves2129}. Axes are in ACS pixels ($0.05 \arcsec /pixel$), and a $20\arcsec$ scale bar is overplotted.}
\label{contours2129}
\end{figure}

\subsection{MACS J2214.9-1359}
\begin{figure}
 \begin{center}
  \includegraphics[width=80mm,trim=0mm 0mm 0mm 0mm,clip]{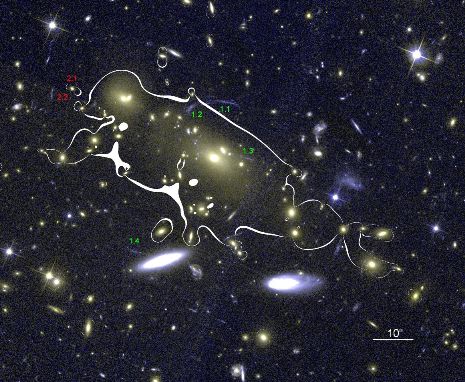}
 \end{center}
\caption{Galaxy cluster MACS J2214.9-1359 ($z=0.503$) imaged
with Hubble/ACS F555W and F814W bands. The critical curve overlaid in white corresponds to system 1, enclosing a critical area of an effective Einstein radius of $\sim140$ kpc at the redshift of the cluster, for an estimated source redshift of $z_{s}\sim2$.}
\label{curves2214}
\end{figure}
\begin{figure}
 \begin{center}
  \includegraphics[width=85mm,trim=0mm 0mm 0mm 0mm,clip]{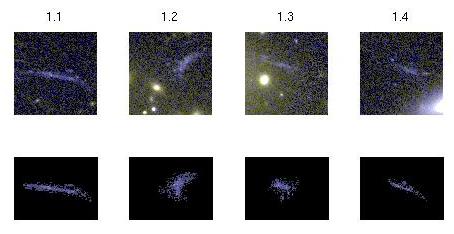}
 \end{center}
\caption{Reproduction of system 1 in MACS J2214.9-1359 by our model, by delensing image 1.1 into the source plane, and relensing the source-plane pixels onto the image plane to accurately form the other images.}
\label{rep2214}
\end{figure}

The galaxy cluster MACS J2214.9-1359 ($z=0.503$) comprises several prominent large blue arcs, 4 of which we match as system 1 (see Figures \ref{curves2214} and \ref{rep2214}). This system might include an additional image, with mirror symmetry to 1.4, if the nearby foreground object is prominently included, yet we do not use this image for constraining our model. We estimate the redshift of systems 1 and 2 as $z_{s}\sim2$, which encloses a mass of $1.25\pm0.10 \times 10^{14} M_{\odot}$ in a critical area with an effective Einstein radius of $\theta_{e}=23\pm2 \arcsec$.
The reference centre of our analysis is fixed at: RA = 22:14:56.59, Dec = -14:00:17.23 (J2000.0), where one arcsecond corresponds to 6.16 kpc at the redshift of this cluster.

\begin{figure}
 \begin{center}
  \includegraphics[width=80mm,trim=5mm 0mm 5mm 5mm,clip]{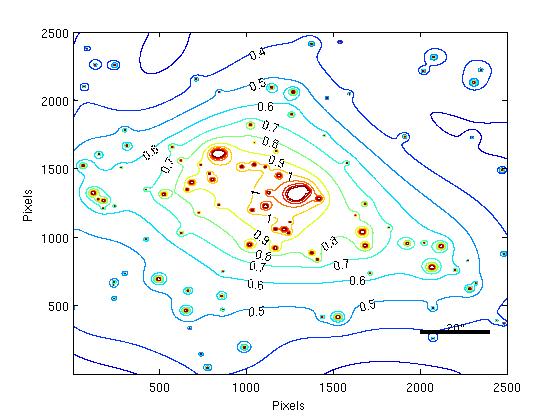}
  \end{center}
\caption{2D surface mass distribution ($\kappa$), in units of the
critical density, of MACS J2214.9-1359. Contours are shown in linear units, derived from
the mass model constrained using the multiply-lensed images seen in
Figure \ref{curves2214}. Axes are in ACS pixels ($0.05 \arcsec /pixel$), and a $20\arcsec$ scale bar is overplotted.}
\label{contours2214}
\end{figure}

\section{Discussion}
Here we examine the SL properties of the whole sample. The Einstein radii, enclosed mass, and high-$z$ magnifications are listed in Table \ref{results_table}, and discussed in the following subsections.
\begin{table*}
\caption{Summary and results of the SL analysis. Part of the following data are based on Ebeling et al. (2007; see also Table \ref{sample}). \emph{Column 3:} Effective Einstein radius, in arcseconds. Simply the square root of the area enclosed within the critical curves divided by $\pi$; \emph{Column 4:} Effective Einstein radius, in kpc, for $z_{s}\sim2$. An uncertainty of $\Delta z\pm0.5$ in the estimated source redshift results in a typical $\sim10\%$ deviation in the lensing distance ratio, and usually up to such deviation also in the Einstein radii. \emph{Column 5:} Mass enclosed within the critical curves, in $10^{14} M_{\odot}$. The quoted errors correspond to the different models. They do not include the source redshift uncertainty where exists. An uncertainty of $\Delta z\sim$0.5 can cause a typical mass uncertainty of $^{+50}_{-10} \%$; \emph{Column 6:} $M/L_{B}$, in $(M/L)_{\odot}$, fluxes were converted to luminosities using the LRG template described
in Ben\'itez et al. (2009); \emph{Column 7:} Lower limit of the highly magnified ($> \times10$) area for a high-$z$ source at $z_{s}\sim8$, in square arcminutes; \emph{Column 8:} image-plane RMS, in arcseconds; for more details see also Table \ref{sample}.}
%\vspace{0.5cm}
\label{results_table}
%\begin{footnotesize}
\begin{center}
\begin{tabular}{|c|c|c|c|c|c|c|c|c|c|c|c|}
  \hline\hline
%  MACS & $z$ & $\theta_{e}$'' & $\simeq r_{e}$ & Mass & $\simeq(M/L_{B})_{\odot}$&
  MACS & $z$ & $\theta_{e}$'' & $\simeq r_{e}$ & Mass & $(M/L_{B})_{\odot}$ & $\mu >10$ & $RMS$ & $\sigma$ & $L_{x}$ & $KT (kev)$ & M.C.E.\\
%$\mu >10$& $RMS$ && $L_{x}$ & $KT (keV)$ & M.C.E. &
%   & & &(kpc)&($10^{14} M_{\odot}$)&&&image & &Chandra&&\\
   & & &(kpc)&($10^{14} M_{\odot}$)&&&image & km s$^{-1}$&(Chandra)&&\\
  \hline
  J0018.5+1626 & 0.545 & $24\pm2$& 154& $1.46\pm0.1$ & 278& $3.1\sq\arcmin$&$1.5\arcsec$ &$1420^{+140}_{-150}$ & $19.6\pm{0.3}$ & $9.4\pm{1.3}$ & 3\\
  J0025.4-1222 & 0.584 & $30\pm2$& 199& $2.42^{+0.10}_{-0.13}$ & 171&$3.1\sq\arcmin$&$2\arcsec$ &$740^{+90}_{-110}$ & $8.8\pm{0.2}$  & $7.1\pm{0.7}$ & 3\\
  J0257.1-2325 & 0.505 & $39\pm2$& 241 & $3.35^{+0.58}_{-0.10}$ & 435&$2.9\sq\arcmin$&$1\arcsec$ &$970^{+200}_{-250}$ & $ 13.7\pm{0.3}$ & $10.5\pm{1.0}$&  2\\
  J0454.1-0300 & 0.538 & $13^{+3}_{-2}$& 83 & $0.41^{+0.03}_{-0.01}$ & 159&$2.6\sq\arcmin$&$2.7\arcsec$ &$1250^{+130}_{-180}$ & $16.8\pm{0.6}$ & $7.5\pm{1.0}$ & 2\\
  J0647.7+7015 & 0.591 & $28\pm2$& 187 & $2.07\pm0.1$ & 256&$2.8\sq\arcmin$&$3\arcsec$ &$900^{+120}_{-180}$ & $15.9\pm{0.4}$ & $11.5\pm{1.0}$&  2\\
  J0717.5+3745 & 0.546 & $55\pm3$& 353 & $7.4\pm0.5$ & 370&$3.3\sq\arcmin$&$2.2\arcsec$ &$1660^{+120 }_{-130}$ & $24.6\pm{0.3}$ & $11.6\pm{0.5}$&  4\\
  J0744.8+3927 & 0.698 & $31\pm2$&222 & $3.1\pm0.1$ & 380& $2.5\sq\arcmin$&$1\arcsec$ & $1110^{+130 }_{-150}$ & $22.9\pm{0.6}$ & $8.1\pm{0.6}$&  2\\
  J0911.2+1746 & 0.505 & $11^{+3}_{-1}$&68 & $0.28^{+0.02}_{-0.01}$ & 101 &$1.4\sq\arcmin$& $1.5\arcsec$ & $1150^{+260 }_{-340}$ & $7.8\pm{0.3}$ & $8.8\pm{0.7}$&  4\\
  J1149.5+2223 & 0.544 & $27\pm3$ & 173& $1.71\pm0.20$ & 190& $2.6\sq\arcmin$ & $1.8\arcsec$&$1840^{+120 }_{-170}$ & $17.6\pm{0.4}$ & $9.1\pm{0.7}$ & 4\\
  J1423.8+2404 & 0.543 & $20\pm2$& 128 & $1.3\pm0.40$ & 194& $0.7\sq\arcmin$& $3\arcsec$ & $1300^{+120}_{-170}$ & $16.5\pm{0.7}$ & $7.0\pm{0.8}$ & 1\\
  J2129.4-0741 & 0.589 & $37\pm2$& 246& $3.4^{+0.6}_{-0.3}$& 314&$2.6\sq\arcmin$& $3.4\arcsec$ &$1400^{+170}_{-200}$ & $15.7\pm{0.4}$ & $8.1\pm{0.7}$ & 3\\
  J2214.9-1359 & 0.503 & $23\pm2$ &142 & $1.25\pm0.10$ & 153&$1.4\sq\arcmin$& $1\arcsec$ & $1300^{+90}_{-100}$ & $14.1\pm{0.3}$ & $8.8\pm{0.7}$ & 2\\
  \hline
\end{tabular}
\end{center}
\end{table*}

\subsection{The Einstein Radius Distribution}\label{er}

\begin{figure}
 \begin{center}
 \includegraphics[width=80mm,trim=0mm 0mm 0mm 0mm,clip]{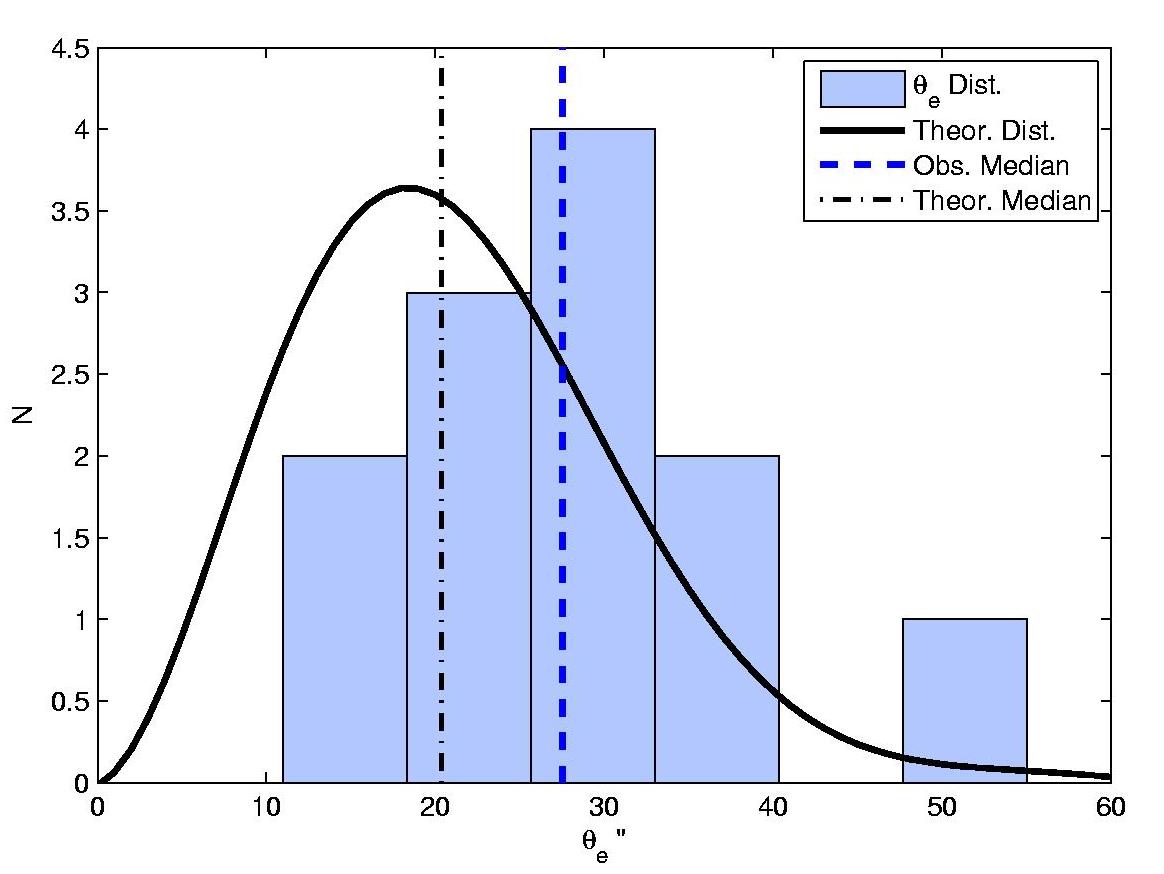}
 \end{center}
\caption{Einstein Radius distribution. This histogram shows the number of clusters per bin, where 6 equally-spaced bins were used to divide the range $11 \arcsec$ to $55 \arcsec$. The solid curve is the expected distribution of Einstein radii as calculate in \S \ref{cdm} for the $\Lambda$CDM model, incorporating both the scatter on the $L_x-M$ relation and the projection bias from lensing. We multiply the resulting theoretical $dN/d\theta_e$ curve by the width of the bins to normalise it. The observed distribution is skewed to larger Einstein radii than predicted. The median values of both the observed and the theoretical distributions are plotted on the histogram in a blue dashed line, and a black dash-dotted line, respectively.}
\label{hist}
\end{figure}

The 12 clusters have effective Einstein radii in the range, 11$\arcsec$ to 55$\arcsec$, with a median value of $27.5\arcsec$, (and a mean value of $28 \pm 3.6\arcsec$) assuming a fixed source redshift of $z_s\simeq2$, see Figure \ref{hist}. This corresponds to a range of physical
radii, 68 kpc to 353 kpc, with a median value of 180 kpc, when transforming to the redshift of each lens. In the hierarchical model large-scale perturbations collapse recently and thus should be found relatively locally.
This sample includes very large Einstein radii, exceeding even the
most impressive lenses previously studied at lower redshifts as analysed by Broadhurst \& Barkana (2008). More recently, Richard et al. (2009b) carefully studied the lensing properties of 20 mainly undisturbed clusters at lower-$z$, selected to have SL features in Hubble data and measured X-ray data, for which their mean Einstein radius is $\theta_{e}=14.45 \arcsec$, a factor of $\sim$two smaller than our sample. This difference reflects
the larger masses of the MACS sub-sample that we studied, which is purely X-ray flux selected and restricted to $z>0.5$. This MACS sample therefore does not suffer from a lensing-selection bias, but the effect of projection bias must be taken into account when evaluating 2D lensing observations with 3D mass profiles predicted by theory (e.g., Hennawi et al. 2007, Oguri \& Blandford 2009, Sereno, Jetzer \& Lubini 2010), as discussed in \S \ref{cdm} below.

\begin{figure}
 \begin{center}
 \includegraphics[width=80mm,trim=0mm 0mm 0mm 0mm,clip]{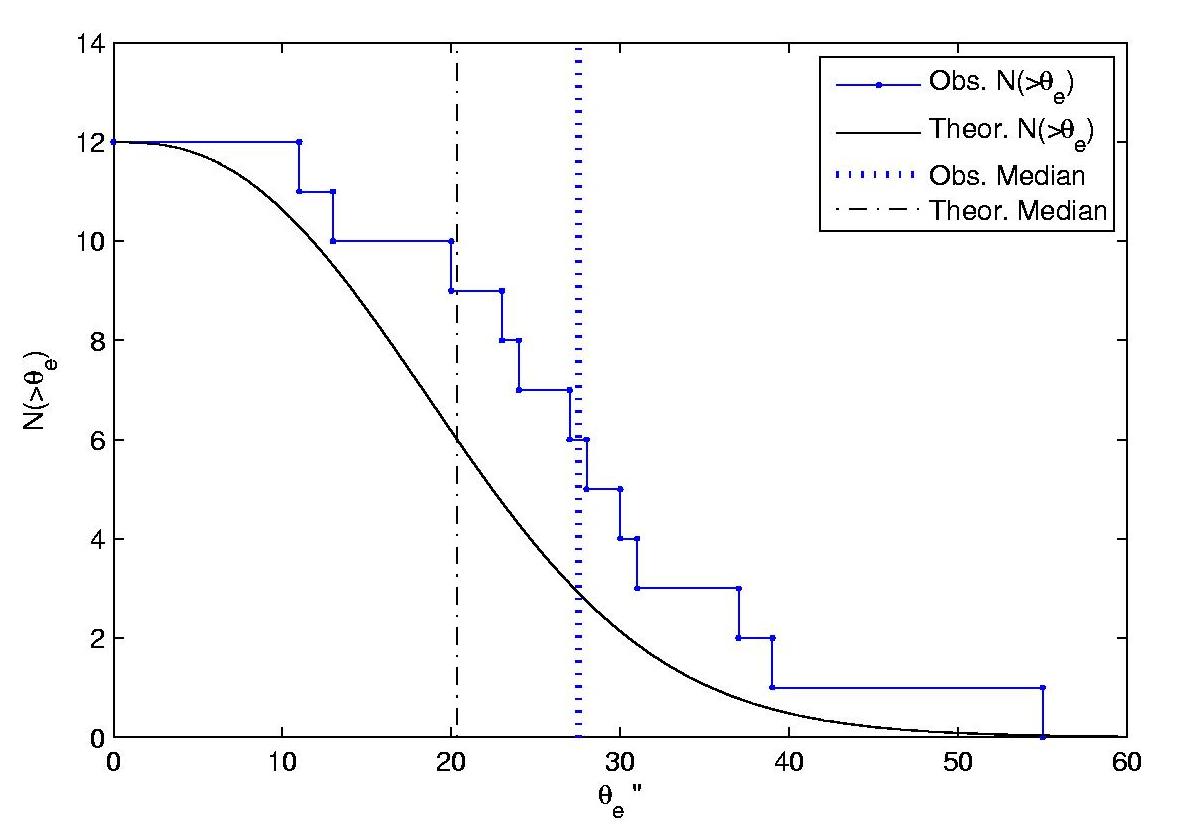}
 \end{center}
\caption{Comparison of the cumulative distribution of observed Einstein Radii (blue solid stairs) and the theoretical distribution predicted by the $\Lambda$CDM model (black solid curve) described in \S \ref{cdm}.The median values of both the observed and the theoretical cumulative distributions are plotted on the histogram in a blue dashed line, and a black dash-dotted line, respectively.}
\label{histR}
\end{figure}

\begin{figure}
 \begin{center}
 \includegraphics[width=80mm,trim=0mm 0mm 0mm 0mm,clip]{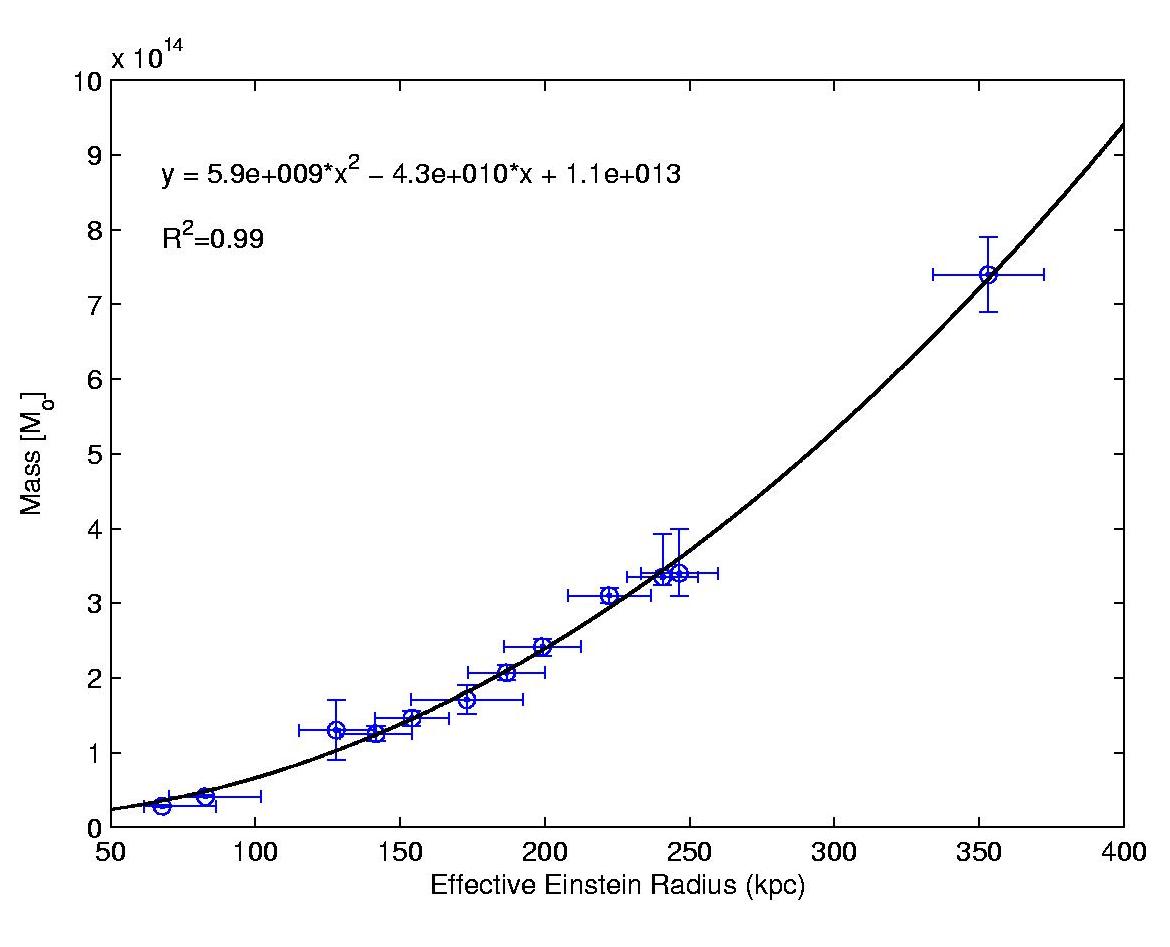}
 \end{center}
\caption{Mass enclosed within the critical curves as a function of the effective Einstein radii. The theoretical relation for symmetric mass distribution, $M \propto r_{e}^{2}$, is tightly followed by the data, indicating that in general the central mass distributions of these clusters are well behaved.}
\label{reme}
\end{figure}

\begin{figure}
 \begin{center}
 \includegraphics[width=80mm,trim=0mm 0mm 0mm 0mm,clip]{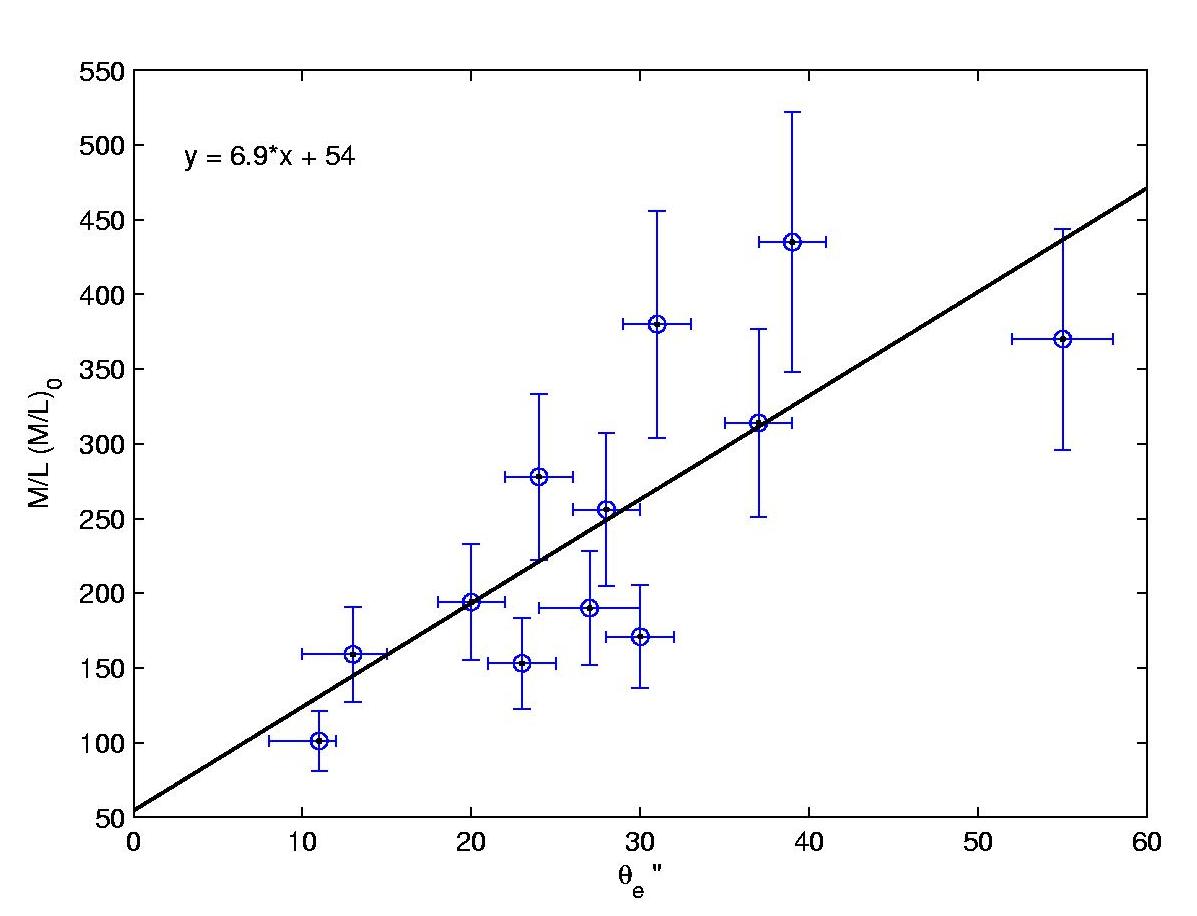}
 \end{center}
\caption{Mass-to-light ratio enclosed within the critical curves as a function of the Einstein radius (in arcseconds).
%The general tendency of $M/L$ to increase with the system scale size is shown in this plot.}
A general trend of increasing $M/L$ with the system scale size is apparent.
}
\label{ml}
\end{figure}

Most of these clusters have resisted analysis by strong lensing despite the long availability of the Hubble data. Our success in ``cracking'' these irregular lenses and defining their critical curves is encouraging for the application of our approach to
 complex unrelaxed clusters more generally. For such clusters, where the deflection fields are not symmetric, the identification of multiple images must be aided by
models that allow for flexibility in describing the mass distribution, as described in section \ref{model}, free of the
 strictures imposed by the use of idealised elliptical potentials in conventional models.

\begin{figure}
 \begin{center}
 \includegraphics[width=80mm,trim=0mm 0mm 0mm 0mm,clip]{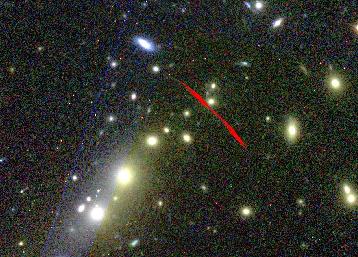}
 \end{center}
\caption{Reproduction of system 1 in MACS J0018.5+1626 by our model. Since the arc is too faint to be noticed here we lens the middle arclet and paint in red the triplet outcome.}
\label{rep0018}
\end{figure}

\subsection{Central Mass Distributions}

The masses enclosed by the critical curves range from $2.8^{+0.2}_{-0.1}\times10^{13} M_{\odot}$ to $7.4\pm0.5\times10^{14} M_{\odot}$ with a median value of $1.9\times10^{14} M_{\odot}$, and a mean value of $2.4^{+0.8}_{-0.2}\times10^{14} M_{\odot}$ as shown in table \ref{results_table}. These
are calculated for each cluster by integrating the surface mass density distribution within the 2D model critical curve, scaled to a source redshift of $z_s\sim2$. We examine the relation between these Einstein masses, $M_{ein}$ enclosed within the critical curves, and the effective Einstein radii derived above. Theoretically this should of course simply scale as $M_{ein} \propto \theta_{e}^{2}$ for symmetric lenses. Here the lenses are not symmetric, but as can be seen in Figure \ref{reme} the quadratic relation is tightly followed, indicating that asymmetry is not very significant in terms of the mass distribution, and furthermore the measured or assumed redshifts of $z\simeq2$ for the relevant systems is a reliable estimate. In fact, we see clearly in that the 2D mass distributions are in general noticeably rounder than the critical curves which are very sensitive to substructure, as can be seen by comparing the 2D mass distributions with the critical curves, shown in Figures \ref{curves0018} - \ref{contours2214}. Further encouragement for the accuracy of SL mass estimates is presented in recent work by Meneghetti et al. (2010), showing that SL methods based on parametric modelling are accurate at the level of few percent at predicting the projected inner mass.

Richard et al. (2009b) found a mean mass of $1.95 \times 10^{14} M_{\odot}$ enclosed mass (within R$<$250 kpc) for their sample of 20 clusters, while Smith et al. (2005) examined the mass distribution of 10 X-ray selected clusters, and found a
%mean similar
similar mean
mass of  $1.86 \times 10^{14} M_{\odot}$ (see also Richard et al. 2009b). These values are in good agreement with our sample median mass ($1.9\times10^{14} M_{\odot}$), and slightly lower than our sample mean mass which is boosted by few extremely massive clusters (e.g., MACS J0717.5+3745).

We examine the behaviour of the central mass-to-light
ratio, $M/L_{B}$, with the Einstein radius and shown in Figure \ref{ml}. The
luminosity is summed over all cluster sequence galaxies identified as described in section \ref{model}. A clear correlation is obtained such that $M/L_{B}$ increases with the Einstein radius, and with values quite typical of other massive clusters (Sarazin 1988, Rines et al. 2004, Medezinski et al. 2010).
This scaling towards higher $M/L_B$ provides confidence in our method. At the high mass end our result is
similar to the peak M/L found in the detailed weak lensing profile studies of Medezinski
et al. (2010). Measuring $M/L_{B}$ in high-$z$ clusters is of interest also for the added insight
on the amount of DM initially associated with individual galaxies as compared
with the overall cluster DM component (see also Sarazin 1988).

\subsection{The Magnification Distribution}

As can be seen in Table \ref{results_table}, most of the sample clusters provide large regions ($\geq2.5 \sq\arcmin$) of highly magnified ($>\times 10$) sky, useful for the search of the first stars and galaxies (e.g., Zackrisson et al. 2010). One has to be cautious when making such a statement since unlike the critical curves and the enclosed mass, the magnification is sensitive to the mass profile which requires source redshift estimation for several sources, which we have here for only a few clusters (section \ref{model}). Still, as explained in \S 2, reasonable constraints are put on the mass profile slope by the image-plane minimisation, which generally has a broad minimum at the same point in the parameter space where the slope is also approximately correct. In addition, we are able to put a lower boundary on the high magnification area, by choosing the steeper profiles which naturally generate lower magnifications, or by assuming a boosted redshift for the main multiply-lensed system decreasing the lensing distance ratio for high-$z$ galaxies. Thus we constrain the area of high magnification for a high-$z$ source at the current limit of $z_{s}\sim8$, showing that most of this sample clusters are excellent targets for high-$z$ galaxy searches due to these large high-magnification areas and lens power. We estimate, based on previously analysed clusters (e.g., Cl 0024+1654, Zitrin et al. 2009b; MACS J1149.5+2223, Zitrin \& Broadhurst 2009; MACS J0717.5+3745, Zitrin et al. 2009a) that in practice the areas of high magnification are about 10-20\% higher than the lower limits presented in Table \ref{results_table}, with an upper limit of about 40-50\%.

\begin{figure}
 \begin{center}
 \includegraphics[width=85mm,trim=0mm 0mm 0mm 0mm,clip]{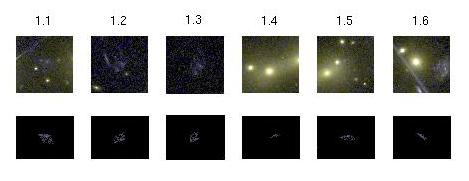}
 \end{center}
\caption{Reproduction of system 1 in MACS J0647.7+7015 by our model, by delensing image 1.1 into the source plane, and relensing the source-plane pixels onto the image plane to accurately form the other images.}
\label{rep0647}
\end{figure}

\begin{figure}
 \begin{center}
 \includegraphics[width=70mm,trim=0mm 0mm 0mm 0mm,clip]{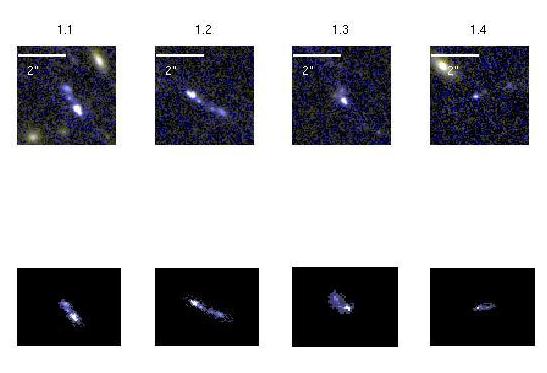}
 \end{center}
\caption{Reproduction of system 1 in MACS J0717.5+3745 by our model. Published originally in Zitrin et al. 2009a.}
\label{rep0717}
\end{figure}

\begin{figure}
 \begin{center}
 \includegraphics[width=70mm,trim=0mm 0mm 0mm 0mm,clip]{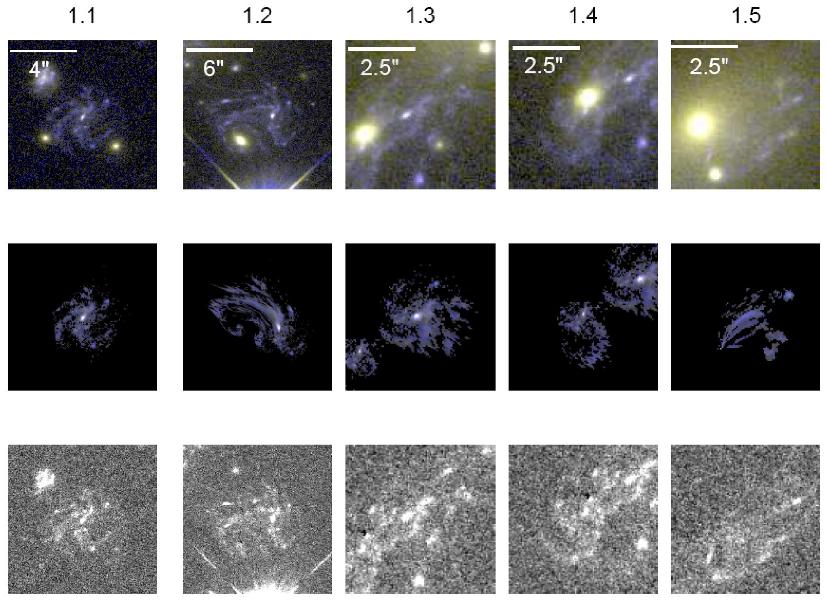}
 \end{center}
\caption{Reproduction of system 1 in MACS J1149.5+2223 by our model. Published originally in Zitrin \& Broadhurst 2009.}
\label{rep11495}
\end{figure}

\subsection{Comparison with $\Lambda$CDM}\label{cdm}

We generally follow Broadhurst \& Barkana (2008; see also Sadeh \& Rephaeli 2008)
in constructing the theoretically predicted distribution of Einstein radii.
In particular, we adopt the NFW parameters measured by Neto et al. (2007) for
simulated halos, although we reduce $c_N$ by $10\%$ according to the results of
Duffy et al. (2008) who used the more recently measured values of the
cosmological parameters. Studies based on large numerical simulations
(Zhao et al. 2003,2009, Gao et al. 2008) have found for massive halos a rather weak
decline of $c_N$ with increasing redshift; we neglect this decline, which renders
our results (in terms of the disagreement with observations) conservative.
We then correct the concentration parameter distribution
using Hennawi et al. (2007) to obtain the effective parameters for the
population of clusters observed in projection. Note though that here
we are considering a sample selected by X-ray flux (not lensing
cross-section), so there is no lensing bias, but there still is the
projection bias; this refers to the fact that when halos are seen in
projection (at randomly distributed angles), a different effective
projected $c_N$ is measured than would be obtained from a spherical
analysis of the halo density profile. In particular, projection
increases the mean $c_N$ and adds substantially to the scatter
compared to the distribution of the spherically-measured
$c_N$. Neto et al. (2007) split their halo sample into two groups
(``relaxed'' and ``unrelaxed'') and analysed the statistics of each
group separately, so we do the same, and in the end combine the two
groups according to their relative numbers in the simulation by
Neto et al. (2007), obtaining our predicted results for the total, combined
population of halos. This can then be compared with the observed
sample, which does not select for relaxed clusters but includes both
relaxed and unrelaxed ones.

In order to approximately simulate a flux-selected sample, we need to convert $L_X$ to halo mass.
We use the power-law relation
from Reiprich \& B\"{o}hringer (1999; with an updated Hubble constant), along
with the observed scatter of $L_X$ for a given mass, which we model as
a lognormal distribution with a typical scatter that corresponds to a
factor of 1.5 . Note that this observed relation is based on
$M_{500}$, while the results from the above simulations refer to
$M_{200}$, so we must include the relation between $M_{500}$ and
$M_{200}$ which is itself a function of $c_N$. An important caveat is
that the observed $L_X-M$ relation was measured for relaxed clusters
(whose mass could be estimated from the X-ray emission assuming
hydrostatic equilibrium). We expect that unrelaxed clusters would tend
to have an unusually high $L_X$ for a given mass (e.g., in a
post-merger phase), which would insert more low-mass clusters within
the sample than we assume, leading to smaller predicted Einstein
angles, increasing the discrepancy between the theory and the
observations (again making our results conservative).

On the other hand, it should be noted that much of the scatter in the $L_X-M$ relation probably arises from cool cores in the center (since it is significantly reduced if the inner core is excised), implying a possible bias towards relaxed and more concentrated systems. However, such a bias seems less probable in our case due to the high X-ray luminosity, the relatively high redshift, and the clear spread-out appearance of most of the clusters analysed here.

The observed properties of each halo (flux, Einstein radius
distribution, etc.) change slowly with redshift, so to simplify the
calculation we approximate the clusters as all lying at a typical
redshift $z = 0.55$. However, the halo abundance changes rapidly with
redshift, so we calculate the integrated halo mass function for all
halos observed at $z>0.5$, in the fraction of the sky corresponding to
the MACS survey (around a quarter of the sky). We use the
Sheth \& Tormen (1999) formula, which accurately fits the halo mass function
(i.e., the distribution of $M_{200}$) measured in simulations.

This halo mass function was convolved with
with the $c_N$ distribution to obtain
the distribution of $M_{500}$, and then add the $L_X-M_{500}$ scatter
in order to obtain the halo mass function corresponding to selection
by a given minimum flux. Since the $L_X-M_{500}$ relation is not
strictly valid for unrelaxed clusters, we only use the power-law and
the scatter from this relation but allow some flexibility in the
normalisation, choosing an effective minimum flux that corresponds to
a predicted abundance of 12 clusters (as in the observed sample). This
is reasonable also since we are interested here in testing the
predicted $\theta_e$ distribution, not in testing the predicted
cluster abundance. Note that the flux selection does not affect the
distribution of the very highest $\theta_e$ values, since all
clusters within the survey
solid angle at $z>0.5$ that can produce the largest $\theta_e$'s will
be well above the flux threshold.

Given our predicted, X-ray selected mass function in terms of
$M_{500}$, we must convolve it with the $\theta_e$ distribution of
each halo of mass $M_{500}$. From the simulations we have the $c_N$
distribution (and thus also the $\theta_e$ and $M_{500}$ distribution)
for a given $M_{200}$, but we can invert this conditional probability
using Bayes' theorem. Note that the $M_{200}$ to $M_{500}$ conversion
should not really include the projection bias that is included in
$\theta_e$, but having two different $c_N$ distributions for each
$M_{200}$ would makes things far more complicated. Our simplification
of using a single $c_N$ distribution is reasonable since the
projection bias typically affects $M_{500}$ (for a given $M_{200}$)
only by $15\%$. Also, since $c_N$ affects both $M_{500}$ and
$\theta_e$ in the same direction, the effect on our final result
(which derives $\theta_e$ for a given $M_{500}$) is much smaller.

Our predicted $\theta_e$ distribution is compared to the observations
in Figure \ref{histR}. We compare the two cumulative distributions
$N(>\theta_e)$, i.e., the total number of clusters expected above
$\theta_e$, where the theoretical distribution has been normalised (as
noted above) to agree with the total of 12 clusters in the sample. The
observed distribution clearly lies at higher Einstein radii compared
with the predicted distribution. One indication of this is the most
discrepant cluster, J0717.5+3745 with $\theta_e = 55 \arcsec$.
The theoretical calculation yields a probability of only $3.4\%$ of
finding such a large $\theta_e$ in this $z>0.5$ cluster sample, giving
a significant (though not extreme) 2-$\sigma$ discrepancy. More
generally, the observed distribution resembles the predicted one
except offset towards higher angles by about a factor of 1.4
(comparing, for example, the two medians). The standard K-S statistic
for comparing two distributions gives a probability of $7.4\%$ that
the observed distribution is drawn from the predicted one. This is
also around a 2-$\sigma$ discrepancy, largely independent of the above
number since the K-S statistic focuses on the middle portion of the
probability distributions and is insensitive to the edges. The
relatively low significance of the K-S discrepancy is an inevitable result of the
Poisson fluctuations expected with a sample of only 12 clusters.

The discrepancy that we find, of Einstein radii that are too high by
about a factor of 1.4, appears smaller than the roughly factor of 2
difference found by Broadhurst \& Barkana (2008). This may be explained at least
partially by the difference between a lensing selected and an X-ray
selected sample. A more robust comparison would be possible with an
X-ray selected cluster sample in which the clusters have reliable
virial masses measured using weak lensing. Another caveat is that we
have compared the observed Einstein radii with the predictions based
on pure dark matter simulations. The possible effect of baryons on the
halo density profile has been difficult to study with simulations, since
hydrodynamic simulations that produce a large effect do so along with
a central baryon concentration that disagrees with observations and is
due to the ``overcooling'' problem in simulations. Analytical models
suggest that even without producing a central baryon concentration,
the complex coupled history of baryon and dark matter accretion may
allow the baryons to significantly affect the final central density
profile, potentially alleviating the discrepancy at least partially
(Barkana \& Loeb 2010). On the other hand, recent simulations suggest that the discrepancy actually increases when AGN feedback is taken into account to overcome the overcooling problem (Duffy et al. 2010).

\subsection{Effects of Uncertainties on the Results}
In this work we determine the Einstein radii and projected masses for
the 12 clusters of the $z>0.5$ MACS sample, and compare these
to simulations. We make use of our well-established modelling
method to identify many sets of multiply-lensed images which
are in turn used to constrain the models, and determine the
Einstein radii and projected masses for $z_{s}\simeq2$.
Typical uncertainties in the source redshift estimates of $\Delta
z\pm0.5$ impose only minor uncertainties on the comparison with $\Lambda$CDM.
Firstly, such an uncertainty does not enable a unique
determination of the profile slope, which we do not attempt to
fully constrain here since both the Einstein radius and the
projected mass are indifferent to the mass profile slope and
are determined by the data. Secondly, overestimating a source
redshift would in practice increase the projected mass and the observed Einstein
radius for a source at $z_s=2$, while
maintaining the relation seen in Figure \ref{reme} and thus
resulting only in growth of the discrepancy from $\Lambda$CDM
simulations which are not dependent on the observed mass. On the other hand, underestimating a source
redshift would in practice decrease the observed Einstein
radius and projected mass for a source at $z_s=2$ by less than
10\%, thus insignificantly influencing the results.

\section{Summary}

The MACS $z>0.5$ sample has proven to be of great value for many very different
studies spanning the X-ray through radio spectrum. We have extracted additional useful information from
this sample by solving the strong lensing for these clusters. Previous
discoveries from this sample include the largest lensed images of a
highly-magnified distant spiral galaxy (MACS J1149.5+2223; Zitrin \&
Broadhurst 2009, Smith et al. 2009), the largest known lens (MACS J0717.5+3745; Zitrin et al. 2009a), lensed sub-mm sources (Takata et al. 2003, Borys et al. 2004, Berciano Alba et al. 2007, 2009), and another ``bullet cluster'' (MACS J0025.4-1222; Brada\v{c} et al. 2008b).

We presented mass models and the corresponding critical curves for the 12 high-$z$ X-ray luminous clusters of the sample via strong-lensing analysis in HST/ACS images.
Several of these clusters have only a few
strongly-lensed images and corresponding low masses and Einstein radii, while
most are very massive and rich
with multiply-lensed arcs due to large critical area and high lens power. The equivalent Einstein radii of this sample range from 11$\arcsec$ to 55$\arcsec$ with a median value of $\simeq28\arcsec$ (similar to the mean value), or 180 kpc. The corresponding masses enclosed within these curves range from $2.8^{+0.2}_{-0.1}\times10^{13} M_{\odot}$ to $7.4\pm0.5\times10^{14} M_{\odot}$ with a median value of $1.9\times10^{14} M_{\odot}$, and a mean value of $2.4^{+0.8}_{-0.2}\times10^{14} M_{\odot}$. We find that the enclosed mass follows tightly the quadratic relation with the equivalent Einstein radius, indicating that the deviation from symmetry is not prominent in these clusters and that our measurements are highly accurate given the measured or assumed redshifts. In addition, we have shown that $M/L_{B}$ increases as expected with the scale size of the system, or Einstein radius.

We compare these results to the predictions of  $\Lambda$CDM taking into account
projection biases and find that these predictions fall short of our measured Einstein radii
by a factor of $\simeq1.4$. In addition, the standard K-S statistic gives a
probability of only $7.4\%$ that the observed distribution is drawn from the
predicted one, corresponding to a $\sim2\sigma$ discrepancy.
It is apparent that the observed Einstein radii and implied high masses are not
likely at high redshifts in the context of the $\Lambda$CDM model
when no lensing-selection biases are involved,
showing again
possible
tension with the prediction of the standard model. Finally, as
a result of their unrelaxed mass distributions, most of these clusters cover a very large
area ($>2.5\sq \arcmin$) of high magnification ($>\times 10$) making them
primary targets for high-$z$ galaxy searches, for which substantial HST time
will shortly be forth-coming.

It should be acknowledged that theoretical predictions may clearly entail further uncertainties than those taken into account here, and a revised and more thorough analysis is needed in order to obtain higher significance results once source redshifts are available. Such comparisons should also be independently applied to other large samples, so that a significant number of clusters are compared to the $\Lambda$CDM model in order to statistically support any claimed discrepancy.

\section*{acknowledgments}
We thank the anonymous reviewer of this work for very useful comments. AZ acknowledges Eran Ofek and Salman Rogers for their publicly available Matlab
scripts used in part of this work. AZ thanks Sharon Sadeh for useful discussions,
and Elinor Medezinski for sharing some Subaru data. This research is being
supported by the Israel Science Foundation (ISF) grant 1400/10. RB is supported by the ISF grant 823/09. ACS was developed under
NASA contract NAS 5-32865. This research is based on observations provided in the
Hubble Legacy Archive which is a collaboration between the Space Telescope Science
Institute (STScI/NASA), the Space Telescope European Coordinating Facility
(ST-ECF/ESA) and the Canadian Astronomy Data Centre (CADC/NRC/CSA).
Part of this work is based on data collected at the Subaru
Telescope, which is operated by the National Astronomical
Society of Japan.

\bsp
\label{lastpage}

\end{document}